\documentclass[preprint,12pt]{elsarticle}

\usepackage{amssymb}
\usepackage{amsmath}
\usepackage[utf8]{inputenc}
\usepackage{setspace}
\usepackage[margin=1.25in]{geometry}
\usepackage{graphicx}
\graphicspath{ {./figures/} }
\usepackage{subcaption}
\usepackage{amsmath}
\usepackage{amsfonts}
\usepackage{lineno}
\usepackage{hyperref}
\usepackage{xcolor}

\journal{Elsevier}

\begin{document}

\begin{frontmatter}



\title{Searching for fixed points of dynamical systems by the Explorative Relaxation Redistribution Method}


\author{Eliodoro Chiavazzo} 

\affiliation{organization={Department of Energy, Politecnico di Torino},
            addressline={Corso Duca Degli Abruzzi}, 
            city={Torino},
            postcode={10129}, 
            country={Italy}}

\begin{abstract}
Being able to effectively locate saddle (and other fixed) points in dynamical systems holds tremendous implications in a number of applications in engineering and science, among which the study of rare events in molecular simulations stands as one of the most prominent field of interest.
Although there is a vast literature on methods aiming at addressing this challenge, open issues still remain in several aspects such as computational efficiency and easy implementation.
In this work, we suggest that the Relaxation Redistribution Method (RRM) - formerly introduced to solve the invariance equation in the context of stiff dynamical systems ruling detailed chemical kinetics - can be reformulated to locate saddle and other fixed points even for stochastic simulators driven by effective energy gradients.
This new formulation of the RRM is referred to as the Explorative Relaxation Redistribution Method (ERRM).
Benchmarks based on the popular Mueller-Brown and other potentials are used to test the ERRM.
\end{abstract}


\end{frontmatter}

\section{INTRODUCTION}
%
Saddle points and transition states are central to understanding rare events and dynamical processes in diverse scientific fields including physics, material engineering, chemistry, and biology. 
Such points are key for analyzing kinetic bottlenecks that govern transitions between metastable states, whether in molecular folding mechanisms \cite{dill2008protein,bernardi2015enhanced,hummer2003coarse,noe2009constructing,faccioliprotein} or the identification of critical transitions in chemical reactions characterized by intricate free-energy landscapes \cite{anslyn2006modern,raucci2022discover}.
Despite their significance, pinpointing and characterizing saddle points in high-dimensional systems remains a formidable challenge, primarily due to the computational intensity required to explore the relevant regions of configuration space.

The literature on rare events and all the associated methods is vast, and the interested reader can find details in many relevant published works \cite{torrie1977nonphysical,sugita1999replica,zhou2006replica,yang2019enhanced,kang2024computing}.

In this context, established techniques like the nudged elastic band (NEB) \cite{NEB01} and string methods \cite{weinan2002string} may rely on preexisting knowledge of both reactant and product states, and often make use of global collective variables to define the reaction path.
Furthermore, enhanced sampling strategies, such as Metadynamics and adaptive biasing force methods \cite{Meta01}, seek to mitigate these issues by extending the exploration of configuration space. 
However, their frequent reliance on predefined reaction coordinates and inefficiency in navigating high-dimensional manifolds can limit their application.

Interestingly, recent progress in data-driven methods and manifold learning has introduced promising alternatives \cite{bonati2021deep,jung2023machine}.
To this end, intrinsic approaches such as the iMapD method \cite{chiavazzo2017PNAS,georgiou2017exploration,faccioliiMapD} and other methods \cite{Clementi1d,Maggioni2011} may provide a convenient way to locate saddle points without prior knowledge of collective variables or exhaustive state-space exploration \cite{bello2023gentlest}.

These methods utilize local sampling and dimensionality reduction tools, such as diffusion maps \cite{coifman2005geometric,chiavazzo2014processes}, to reveal the manifold's low-dimensional structure.
Adaptive frameworks employing these techniques have demonstrated effectiveness in systematically advancing from stable equilibria to saddle points by iteratively sampling and resolving local manifold charts.

Expanding on these advances, we propose an innovative approach integrating enhanced sampling, manifold learning, and adaptive exploration to identify saddle points in dynamical systems. 
Our methodology addresses two core challenges: (i) the inherent difficulty of identifying kinetically significant transition states in high-dimensional spaces and (ii) the absence of prior knowledge of optimal reaction coordinates. 
Applications to canonical toy problems, including the Müller-Brown potential and Lennard-Jones interacting atoms in a plane, showcase the method's effectiveness and adaptability.
This is achieved by leveraging the Relaxation Redistribution Method (RRM).


The RRM method has been formulated in both local and global implementations \cite{chiavazzo2011adaptive,chiavazzo2009PhDThesis,chiavazzo2012approximation,kooshkbaghi2014global}.
Although introduced within a completely different context, the Relaxation Redistribution Method (RRM) can be regarded as a dynamic generalization to any dimensions of what - in the field of rare events studies - is known as the string method by Vanden-Eijnden and collaborators \cite{weinan2002string,maragliano2006string}.
Furthermore, the Explorative RRM (ERRM) to be discussed here takes inspiration from the {\it equation free} approach by Kevrekidis and collaborators as it aims at a speedup of a complex dynamics by exploiting slow-fast decomposition and smoothness at a Slow Invariant Manifold (SIM) embedded in an ambient phase space. \cite{frewen2009exploration,gear2003equation,kevrekidis2004equation,chiavazzo2014processes}.

This manuscript is organized in sections as follows.
In Section \ref{RRMreview}, the Relaxation Redistribution Method is first briefly reviewed and afterwards extended to a second-order approximation.
In Sections \ref{surfingsection} and \ref{secnavigation}, the several steps of the explorative RRM (ERRM) are defined, with the mathematical details of the adopted methods described in Section \ref{MaterialsMethod}.
Results on selected test cases are presented in Section \ref{results}.
Finally, conclusions and discussions are presented in Section \ref{conclusions}.

\section{The Relaxation Redistribution Method}\label{RRMreview}
\begin{figure}[h]
    \centering
        \includegraphics[width=0.97\textwidth, height=2.5in]{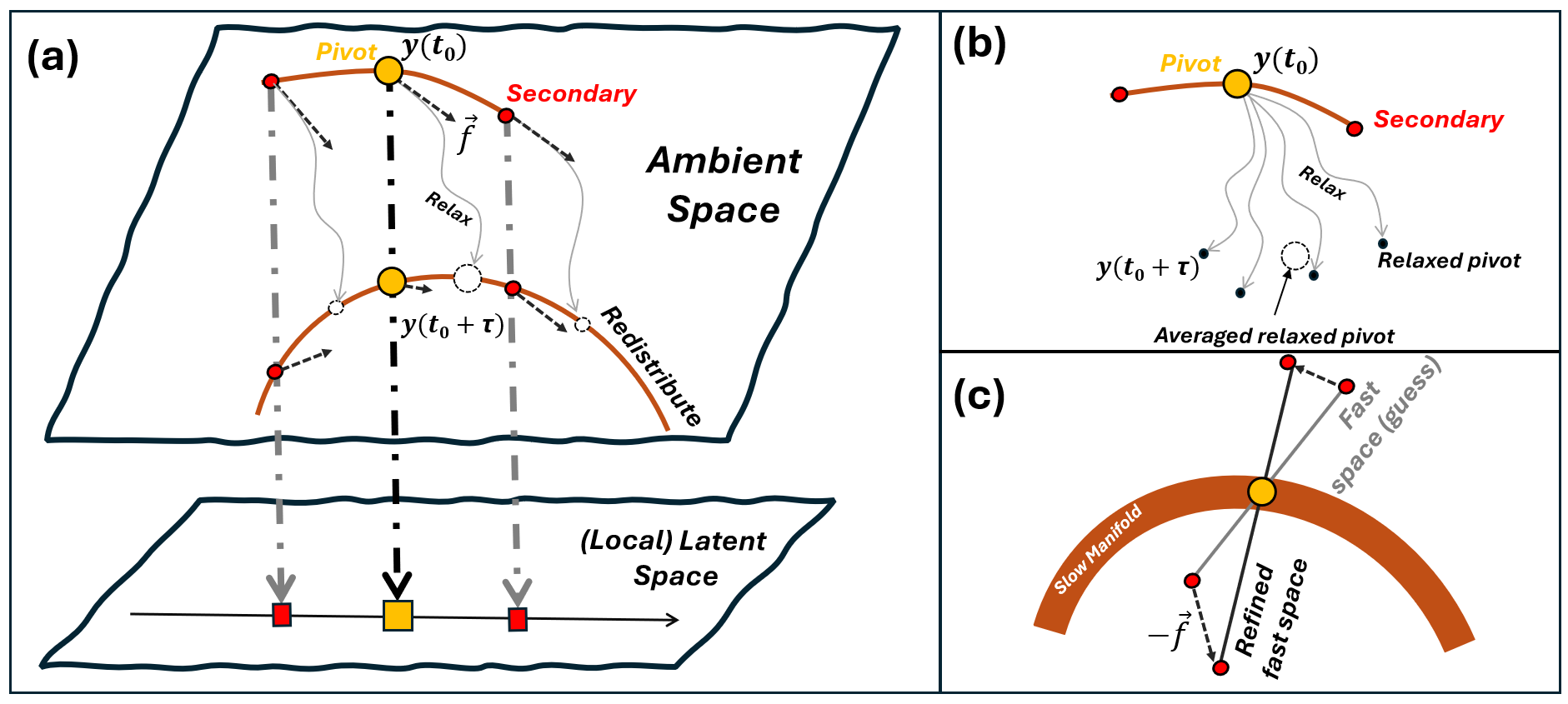}
        \caption{Pictorial representation of the main idea behind the local formulation of the RRM method. (a) Evolution of a small patch towards the slow invariant manifold. For the sake of illustration, we assume a one-dimensional SIM hence, for a second-order local approximation, we search for local description of the SIM including one pivot and two or more secondary points. (b) If the dynamical evolution is dictated by Stochastic Differential Equations (SDEs), multiple short relaxations can be executed from each point with the redistribution step performed on the averaged end points of such short trajectories. (c) Refinement of the local fast space is obtained by relaxing points of an affine space transversal to the manifold under the action of an antiparallel flow field $-\vec{\mathbf{f}}$. See text and Ref. \cite{chiavazzo2012approximation} for details. \label{fig1}}
\end{figure}

The RRM \cite{chiavazzo2011adaptive,chiavazzo2012approximation} in its local formulation rules the evolution of small patches (such as simplices of a properly chosen dimension $d$) towards a Slow Invariant Manifold (SIM) under the action of a flow field $\vec{\mathbf{f}}$:
\begin{equation}\label{sistemadinamicogenerale}
    \frac{d\textbf{y}}{dt}=\vec{\mathbf{f}}(\textbf{y}).
\end{equation}
In the context of model reduction methods for simplifying detailed chemical kinetics \cite{DirkPaper,lebiedz2004computing,ILDM,ValoraniPaper}, the RRM has been first introduced by demonstrating that it searches for stable stationary solutions of the {\it film equation} \cite{GorbanBook} of the dynamics (\ref{sistemadinamicogenerale}).
While full details on the method can be found in previous work \cite{chiavazzo2011adaptive,chiavazzo2012approximation}, it suffices to mention that so far only the linear approximation of the RRM method has been developed.
Nonetheless, for the purpose of this work, higher-order approximations are desirable. 
Hence, as a preliminary step, we further develop below a second-order approximation of the RRM method, whose main idea is reported in Figure \ref{fig1}.
For the sake of completeness, it is important to recognize that the RRM can be applied not only to Ordinary Differential Equations (ODEs) (\ref{sistemadinamicogenerale}), but also to Stochastic Differential Equations (SDEs) as pictorially represented in Figure \ref{fig1} (b).
\subsection{Mapping from the ambient to the latent space and \it{vice-versa}}\label{mappingsec}
One of the key aspects of several model reduction methods is the ability to establish a connection between the immediately accessible high-dimensional space (also referred to as \textit{ambient space}, where the full dynamics happen) with a low-dimensional latent space \cite{chiavazzo2007comparison,chiavazzo2014processes}.
To this end, we assume that - at least locally - a generic point in the full (ambient) space can be projected into a low-dimensional latent space (of dimension $d<<n$). 
Such a projection is anything but unique, and it can be realized by several approaches.
Here, for the sake of simplicity, we adopt the following:
\begin{equation}\label{projectionontomanifold}
    \mathbf{x_0}=\mathbf{D y_0}
\end{equation}
with $\mathbf{x_0}$, $\mathbf{y_0}$ and $\mathbf{D}$ being the $[d \times 1]$ point in the latent space, the $[n \times 1]$ point in the ambient space and the $[d \times n]$ projection matrix, respectively. 
Clearly, the knowledge of the matrix $\mathbf{D}$ allows one to define the latent space locally. 
Although for curved manifolds $\mathbf{D}$ is not fixed, equation (\ref{projectionontomanifold}) provides a rather convenient and straightforward approach for mapping points from the ambient to the latent space (assuming that $\mathbf{D}$ can be estimated and updated at each point of the SIM).

On the other hand, the inverse mapping (i.e. from the local latent space to the ambient space) requires more care.
To this end, following the local formulation of the RRM method, we assume that a pivot point along with a prescribed number of secondary points is available in the ambient space.
More specifically, in the RRM, the number of secondary points depends on both the dimensionality of the SIM (i.e. $d$) and the chosen order of approximation.
In first order approximations (i.e. the pivot and the secondary points forming a simplex), the number of secondary points is dictated by the manifold dimension: $N_s \ge d$.
In second order approximations of interest here, the number of secondary points further increases: 
\begin{equation}\label{Nspoints}
    N_s \ge \frac{d   (d+1)}{2}+d,
\end{equation}
with the adopted equation being reported below in Einstein's notation:
\begin{equation}\label{tensoriallift}
    y^j=A_{jik} x_i x_k +B_{ji} x_i + c_j
\end{equation}
The tensors $\textbf{A}=\{A_{jik}\}$, $\textbf{B}=\{B_{ji}\}$ and $\textbf{c}=\{c_j\}$ are computed as detailed in Section \ref{computetensors} after each RRM iteration on the basis of the ambient and latent coordinates of the pivot and secondary points.
Note that the superscript in $y^j$ of equation (\ref{tensoriallift}) identifies the $j$-th coordinate and does not represent a power.

\subsection{RRM: Refinement of the slow subspace}\label{slowRRMalg}
Endowed with the two mappings (\ref{projectionontomanifold}) and (\ref{tensoriallift}), the RRM aims at finding a local approximation of the SIM starting from an initial set of points comprising one pivot and the associated secondary points.
In general, the initial set of points is expected not to be on the SIM.
As depicted in Fig. \ref{fig1} (a), the RRM consists of fictitious dynamics aiming at driving the pivot and the secondary nodes towards the SIM, by alternatively performing the following two steps:
\begin{itemize}
    \item Relaxation of all points under the forward unbiased dynamics of the considered system. 
    This is particularly convenient as it can be performed even when we only have access to a black-box simulator (i.e. no access to the analytical expression of the right hand-side of (\ref{sistemadinamicogenerale})). 
    The relaxation time $\tau$ is a fixed parameter of the method and is chosen in the order of the characteristic time of the fastest processes of the system \cite{chiavazzo2012approximation}. 
    As depicted in Fig. \ref{fig1} (b), in case of Stochastic Differential Equations (SDEs), the relaxation step of the RRM can be performed by averaging multiple relaxed states starting from the same point.  
    \item Upon completion of the above step, the latent coordinates of the relaxed points will typically differ from the ones corresponding to the starting points.
    For implementing the redistribution step, we first use equation (\ref{tensoriallift}) to best-fit the relaxed points (thus finding the local tensors $\textbf{A}$, $\textbf{B}$ and $\textbf{c}$ by the procedure described in Section \ref{MaterialsMethod}). 
    Subsequently, the relaxed end points will be substituted by those nodes laying on such second order approximation of the SIM and having the same latent coordinates as the starting points.
    This process effectively counteracts the so called \textit{shagreen effect} \cite{chiavazzo2011adaptive}.
\end{itemize}
The RRM continues until a stationary condition is reached: ideally, the relaxation step is fully counteracted by the subsequent redistribution step.
In practice, the stationary condition of the RRM procedure can be detected by imposing that a norm of the overall movement of the pivot after both the relaxation and redistribution steps, divided by a norm of the relaxation movement only, is sufficiently small with respect to a fixed threshold.
In fact, the latter ratio corresponds to imposing that a measure of the \textit{defect of invariance} has to vanish (up to a certain accuracy) on the SIM \cite{chiavazzo2007comparison}. 

\subsection{RRM: Refinement of the fast subspace}
Once the pivot has reached the SIM, if needed, the RRM can also be used to estimate the local fast subspace.
To this end, as done in previous works, we target a linear approximation.
For the sake of simplicity, here we only briefly review the basic idea by referring to the pictorial representation of Figure \ref{fig1} (c), where the fast subspace is one-dimensional.
Generalization to any dimensions can be found in Ref. \cite{chiavazzo2012approximation}.
First, a hypothetical fast subspace is chosen, which is an affine space locally transversal to the SIM.
On that subspace, off-manifold points that align with the pivot can be selected in its neighborhood and relaxed (for a time in the order of $\tau$) under the action of the anti-parallel field $-\vec{\mathbf{f}}$.
The new refined fast local direction is now defined by the pivot and the relaxed points. 
The process can be iterated till a stationary condition is reached (i.e. the anti-parallel movements of the off-manifold points occur in the fast subspace).
Further details are beyond the scope of this work, and the interested reader is referred to Refs \cite{chiavazzo2012approximation,chiavazzo2011adaptive} for deeper discussions.

\subsubsection{What is the fast subspace useful for?}\label{whyfast}
The knowledge of the local fast subspace provides one possible natural parametrization of the SIM in a neighborhood of the pivot.
In other words, a vector basis spanning the null space of the fast subspace can be used to form the $d$ rows of the $\textbf{D}$ matrix in (\ref{projectionontomanifold}).
It is also worth stressing that this is not necessary as the local manifold parametrization is not unique and so is the matrix $\textbf{D}$.
For our purposes, another convenient manifold parametrization around the pivot can be constructed by a vector basis spanning the tangent space of (\ref{tensoriallift}) at the pivot point.  
For implementing the ERRM described below, both options are acceptable.
Needless to say that those are only two possible local parametrizations, with other options also being possible as far as a one-to-one mapping (\ref{projectionontomanifold}) is ensured.

\section{\textit{Surfing} on the Manifold via the RRM}\label{surfingsection}
The approach below takes inspiration from the equation-free framework of Kevrekidis and co-workers \cite{frewen2009exploration,kevrekidis2004equation}.
In particular, in the spirit of the iMapD method \cite{chiavazzo2017PNAS,georgiou2017exploration}, we aim at coping with situations where at least one stable fixed point (e.g. stable configuration of a protein) is observed, and the system is dynamically trapped within a neighborhood of such fixed point, corresponding to a local minimum of the effective free energy.
In this work, we show that the RRM method can be a key tool to force the dynamical system to {\it slide on} the SIM (i.e. the support of the effective free energy), possibly towards other stable fixed points, ideally passing by unstable fixed (saddle) points and revealing reaction transition paths.
Interestingly, the suggested procedure can be implemented without an \textit{a priori} knowledge of the latent space coordinates (sometimes also referred to as \textit{collective variables}), that can be gradually discovered during the process itself.
\begin{figure}[h]
    \centering
        \includegraphics[width=0.95\textwidth, height=3in]{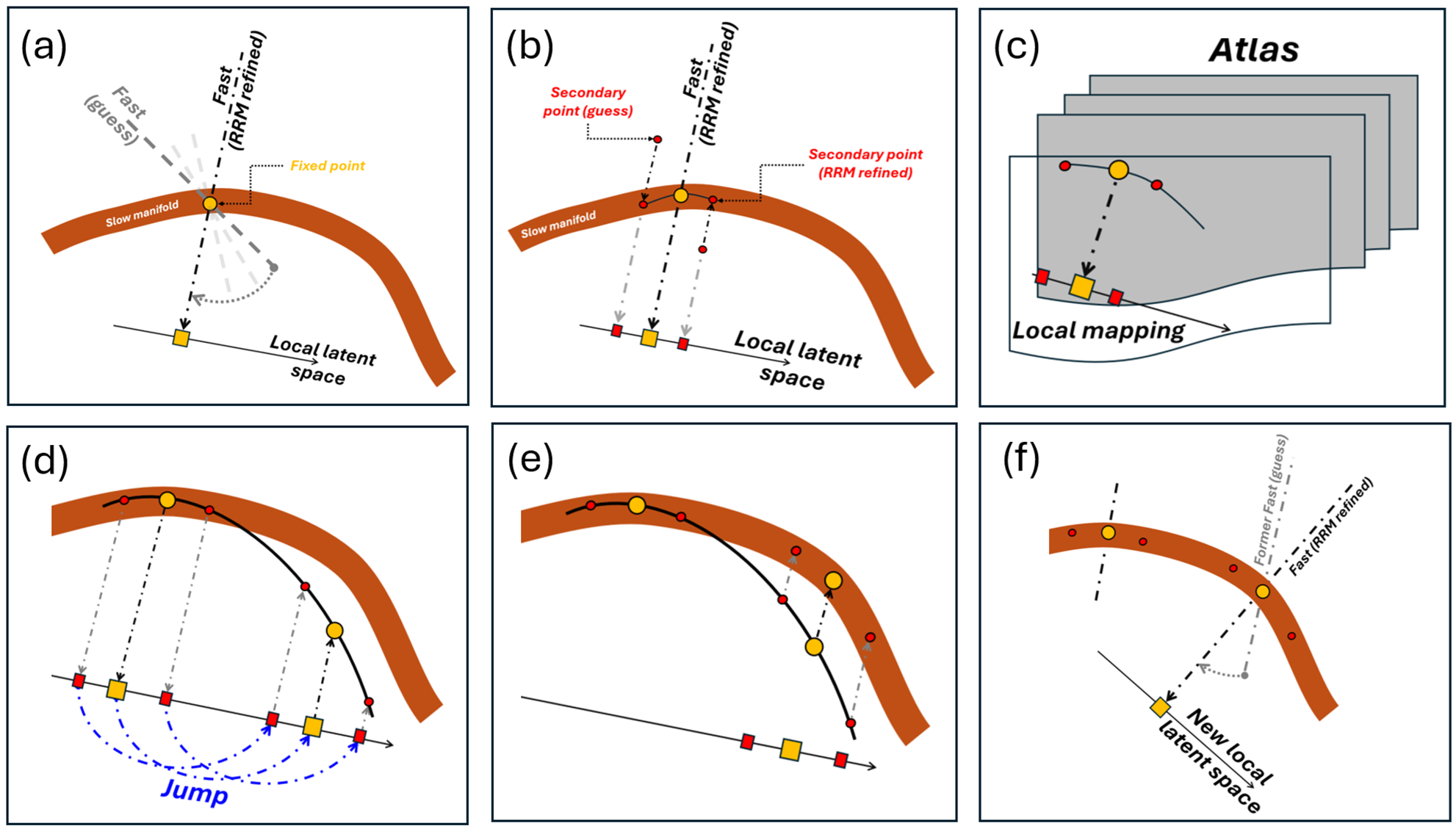}
        \caption{Pictorial representation on the use of the local RRM method for {\it surfing} on the slow manifold. See the text for a detailed description.\label{fig2}}
\end{figure}
The latter navigation process based on the RRM method consists of the following six main steps as schematically reported in Figure \ref{fig2};
\begin{itemize}
    \item (a) Starting from the known fixed point (pivot), we extract a proper local parametrization by refining a fast hypothetic subspace and computing its null space. Again, this is only one possible approach. Another option would be to approximate the local tangent space (see also Section \ref{whyfast});
    \item (b) The local parametrization around the pivot enables one to refine and extract a local second order mapping in the form of (\ref{tensoriallift});
    \item (c) The parameterization matrices \textbf{D} (and the corresponding pivots) are saved so that a global atlas is constructed consistently as described below in Section \ref{secatlasconst}. Strictly speaking, this step can be convenient but not mandatory, especially when $d=1$ and the purpose reduces to a search of unknown fixed points;
    \item (d) Leveraging on the refined mapping for lifting, both the pivot and the secondary points can be moved further away in the latent space towards unexplored regions and the corresponding configurations reconstructed in the ambient space. Owing to the second order accuracy of (\ref{tensoriallift}), we expect that the new reconstructed configurations are located in a close neighborhood of the SIM (for sufficiently short jumps).
    \item (e) The RRM is applied again to correct slight deviations from the SIM. For the sake of computational efficiency, this step can be also substituted (totally or partially) by a relaxation of the unbiased dynamics (i.e. without redistribution step);
    \item (f) The fast subspace is refined again to update the local parametrization of the manifold thus taking into account local curvature (alternatively the local tangent space at the pivot can be used). The refined parametrization matrix $\textbf{D}$ is passed to the Atlas to keep constructing the global mapping, hence the algorithm continues from item (c).
\end{itemize}
\section{Navigation towards other fixed points}\label{secnavigation}
The method described in the above section (\ref{surfingsection}) allows the dynamical system to visit regions of the phase space that otherwise would be very unlikely to be reached during the unbiased simulation.
Clearly, a key aspect is related to the direction along which the jump depicted in Figure \ref{fig2} (d) occurs.
One possible option is the exhaustive search of the iMapD \cite{chiavazzo2017PNAS,georgiou2017exploration}, where the entire set of the boundary points surrounding the available point cloud is outwardly expanded.
While the latter navigation strategy is certainly effective, it may reveal not computationally efficient, especially when the dimension of the SIM is not low (e.g. $d>2$). 
It is therefore desirable to integrate the surfing method of Section  (\ref{surfingsection}) with a smart and efficient navigation strategy capable of driving the exploration towards other fixed points (i.e. saddles or other minima) with the least number of iterations.
In our method, it is worth distinguishing two different cases: i) $d=1$; ii) $d>1$.
Regardless of the intrinsic dimension of the support of the effective free energy surface, the procedure of Figure \ref{fig2} can be performed by imposing $d=1$. 
The latter is certainly the computationally most efficient scenario as there is no ambiguity on the selection of the exploration direction.
When such a choice is successful, we demonstrate below that our method may conveniently and gradually reveal the reaction transition path.
On the other hand, when $d>1$, the smart navigation of the SIM requires a more sophisticated strategy. 
In the latter scenario, here we explore one possible approach based on a preliminary definition of a proper objective function, whose optimization is gradually performed during the navigation by means of active learning algorithms such as those based on Gaussian processes or random forest regressors \cite{bonke2024multi,trezza2022minimal}.
The above concepts are illustrated by means of test cases in Section \ref{results}. 

\section{Materials and methods}\label{MaterialsMethod}

\subsection{Computing tensors of the mapping from latent to the ambient space}\label{computetensors}

The mapping (also referred to as \textit{lift} from the low-dimensional latent space to the ambient space) of points $\mathbf{x} \in \mathbb{R}^d$ to points $\mathbf{y} \in \mathbb{R}^n$ is accomplished by a multivariate quadratic regression. 
To this end, the mapping reported in eq. (\ref{tensoriallift}) can be recast in the following quadratic form for each coordinate $j$ of an ambient space point $\textbf{y}=\{y^1,...,y^n\}$:
\begin{equation}\label{vectoriallift}
    y^j = \mathbf{x}^\top \mathbf{A}_j \mathbf{x} + \mathbf{B}_j \mathbf{x} + c_j, \quad j = 1, \dots, n,
\end{equation}
where $\mathbf{A}_j \in \mathbb{R}^{d \times d}$ is the matrix of quadratic coefficients, $\mathbf{B}_j \in \mathbb{R}^d$ is the vector of linear coefficients, and $c_j \in \mathbb{R}$ is a constant term.
The aim is to compute the following items:
\begin{itemize}
    \item $\mathbf{A} \in \mathbb{R}^{n \times d \times d}$: a tensor containing the $n$ matrices $\mathbf{A}_j$ of quadratic coefficients for each output dimension.
    \item $\mathbf{B} \in \mathbb{R}^{n \times d}$: a matrix containing the $n$ vectors $\mathbf{B}_j$ of linear coefficients.
    \item $\mathbf{c} \in \mathbb{R}^n$: a vector of $n$ constant terms $c_j$.
\end{itemize}

To that end, we construct an auxiliary matrix $\mathbf{H} \in \mathbb{R}^{m \times (d^2 + d + 1)}$ (with $m=N_s+1$), where each row represents a vector consisting of:
\begin{enumerate}
    \item All $d^2$ quadratic terms $x_i x_k$ for $i, k = 1, \dots, d$; 
    \item All $d$ linear terms $x_i$ for $i = 1, \dots, d$;
    \item A constant term $1$.
\end{enumerate}
Hence, we find the least-squares solution by solving the linear system $\mathbf{H} \boldsymbol{\theta}_j = \mathbf{y}^j$, where $\boldsymbol{\theta}_j$ is a vector containing the coefficients of the quadratic, linear, and constant terms for the $j$-th output dimension. 
This can be performed independently for each output dimension $j = 1, \dots, n$.

From the solution $\boldsymbol{\theta}_j$, the coefficients are extracted as follows: i) The first $d^2$ elements of $\boldsymbol{\theta}_j$ are reshaped into the $d \times d$ matrix $\mathbf{A}_j$ (quadratic coefficients); ii) The next $d$ elements form the vector $\mathbf{B}_j$ (linear coefficients); iii) the last element is the scalar $c_j$ (constant term).
%
These coefficients are finally stored in the corresponding output arrays $\mathbf{A}$, $\mathbf{B}$, and $\mathbf{c}$.

\subsection{Atlas: Making local coordinates global}\label{secatlasconst}
In the proposed approach, we gradually have access to manifold points. 
As those points are located on a low dimensional SIM, well established non-linear manifold learning methods (e.g. DMAPS or ISOMAP \cite{coifman2005geometric,isomap2002}) can be safely used to establish a global parameterization.
Nonetheless, considering that during the ERRM we gradually have access to local mappings (via the matrices $\mathbf{D}$ associated to the corresponding SIM points), it appears natural to attempt to leverage on them for constructing a global mapping from the high-dimensional ambient space to a low-dimensional latent space.
We describe below one possible approach to conveniently combine individual local mappings into a coherent global atlas by propagating local information across a graph of data points, thus aiming at a unified representation of the latent space.

\subsubsection{Graph Construction}

Given a set of points $\{\mathbf{c}_1, \mathbf{c}_2, \dots, \mathbf{c}_M\}$ in the high-dimensional space and the associated projection matrices $\{\mathbf{D}_1, \mathbf{D}_2, \dots, \mathbf{D}_M\}$ (properly aligned as discussed in the Section \ref{alignDsec} below), we first construct a graph where nodes represent the points and edges connect each point to its nearest neighbors. 
The nearest neighbors are determined using the $k$-nearest neighbors (kNN) algorithm, with $k$ being a user-defined or automatically chosen parameter. To ensure efficient propagation of local mappings, self-connections in the kNN graph are removed, resulting in a graph where each point is connected only to its neighbors.

It is worth noticing the underpinning hypothesis that we are making below, namely the accessible points homogeneously sample the SIM region of interest so that - upon a proper choice of the above parameter $k$ - the graph correctly capture the manifold topological structure. 
Clearly, when this is not valid, the reported procedure can fail.
Nonetheless, without a loss of generality, the ERRM can be performed by resorting to established and robust approaches for manifold learning such as Diffusion Maps (DMAPs).

\subsubsection{Initialization and Propagation}

The propagation begins from a reference point, $\mathbf{c}_{\text{ref}}$. 
The global coordinate of the reference point is initialized as $\bar{\mathbf{x}}_{\text{ref}} = \mathbf{0}$. 
For the purpose of this work, $\mathbf{c}_{\text{ref}}$ might be the starting stable fixed configuration or a nearby point.
A breadth-first traversal (BFT) is then performed on the graph.
BFT is a notorious graph exploration algorithm that systematically visits nodes in increasing order of distance from the starting point. %
BFT expands outward level by level, thus ensuring that each point in the manifold is assigned a global coordinate in an order that respects the shortest-path structure of the nearest-neighbor graph.
The BFT is implemented using a queue data structure, following these steps:

\textbf{Initialization:} i) Mark all points as unvisited; ii) Initialize a queue and place the reference point $\mathbf{c}_{\text{ref}}$ in it; iii) Set the global coordinate of the reference point: $\bar{\mathbf{x}}_{\text{ref}} = \mathbf{0}$.

\textbf{Propagation through the queue:} i) Extract the current point $\mathbf{c}_i$ from the front of the queue; ii) Identify its nearest neighbors that have not been visited; iii) For each unvisited neighbor $\mathbf{c}_j$ we compute the displacement in the high-dimensional space:
            \begin{equation}
                \Delta \mathbf{c}_{ij} = \mathbf{c}_j - \mathbf{c}_i.
            \end{equation}
project the displacement onto the local latent space using the local projection matrix $\mathbf{D}_i$:

            \begin{equation}
                \Delta \bar{\mathbf{x}}_{ij} = \Delta \mathbf{c}_{ij} \mathbf{D}_i^\top.
            \end{equation}
update the global coordinate of $\mathbf{c}_j$:
            \begin{equation}
                \bar{\mathbf{x}}_j = \bar{\mathbf{x}}_i + \Delta \bar{\mathbf{x}}_{ij}.
            \end{equation}
 and finally mark $\mathbf{c}_j$ as visited and add it to the queue.

\textbf{Termination:} The process terminates when the queue is empty, meaning all reachable points have been assigned global coordinates.

\begin{figure}[h]
    \centering
    \includegraphics[width=0.45\textwidth]{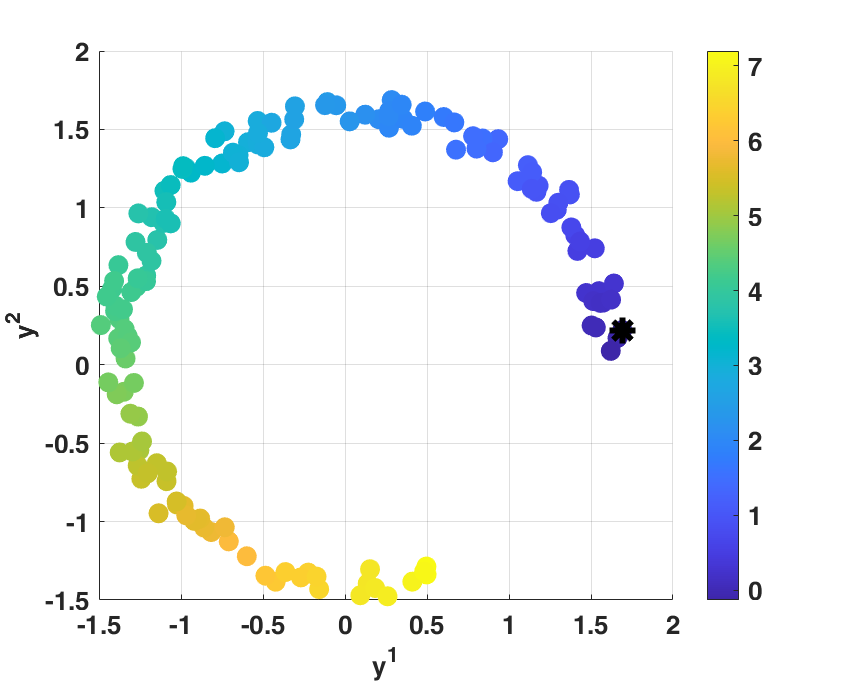}
    \includegraphics[width=0.45\textwidth]{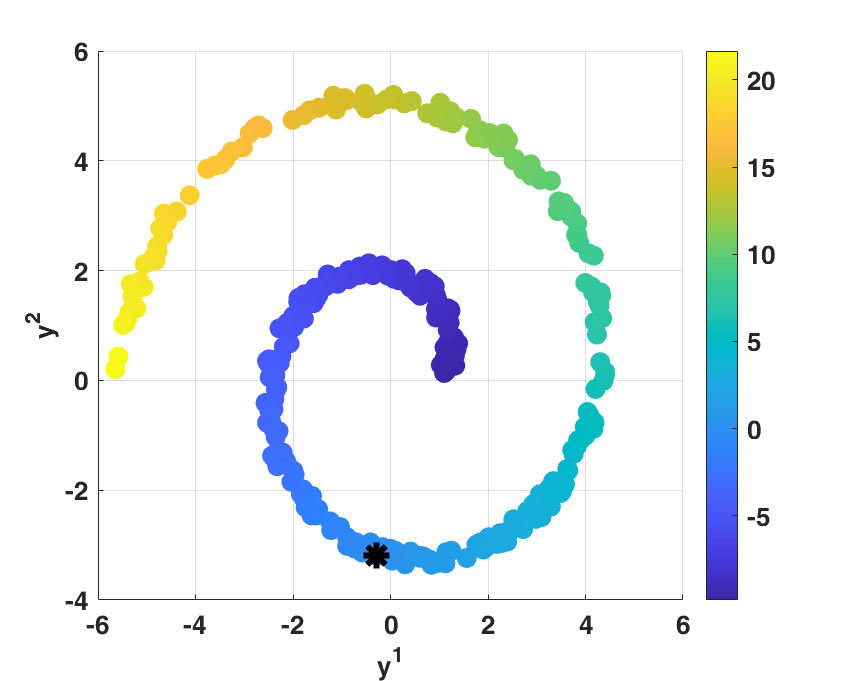}
    \caption{Computation of the one dimensional global coordinate using the algorithm described in Section \ref{secatlasconst}. Left-hand side: 150 randomly chosen points lay on a circle of radius $r=1.5$. Noise is introduced in each coordinate by adding uniformly randomly samples from the interval $[0, 0.2]$. Right-hand side: 350 randomly chosen points on an Archimedes spiral. The black star denotes the reference point where the latent coordinates vanish. Noise is introduced in each coordinate by adding uniformly randomly samples from the interval $[0, 0.3]$. In both cases, matrices $\mathbf{D}$ are constructed on the basis of the local tangent unit vector.}
    \label{Fig1DatlasCirle}
\end{figure}
\begin{figure}[h]
    \centering
    \includegraphics[width=0.45\textwidth]{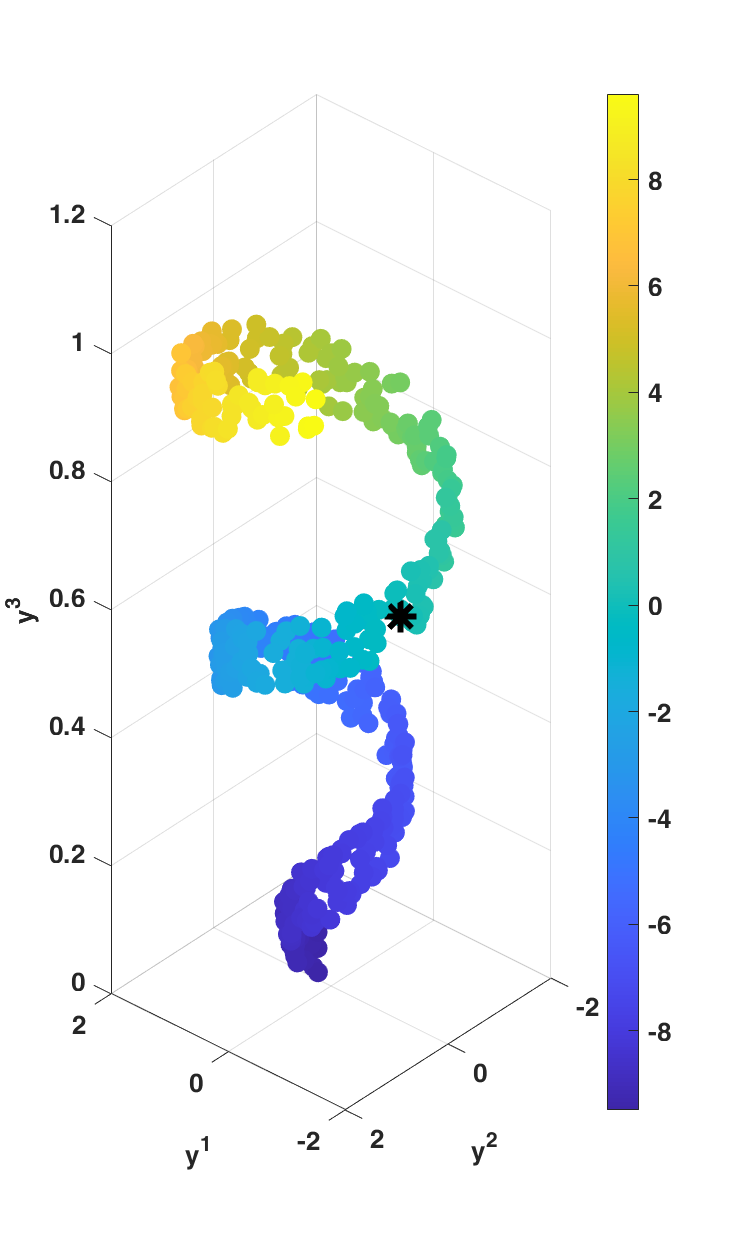}
    \caption{500 randomly chosen points on a Fermat spiral with $\vartheta_{max}=\frac{3}{2} \pi$. The black star denotes the reference point where the latent coordinates vanish. Noise is introduced in each coordinate by adding uniformly randomly samples from the interval $[0, 0.1]$. Matrices $\mathbf{D}$ are constructed on the basis of the local tangent unit vector.}
    \label{Fig1DatlasFermat}
\end{figure}

\begin{figure}[h]
    \centering
    \includegraphics[width=0.98\textwidth]{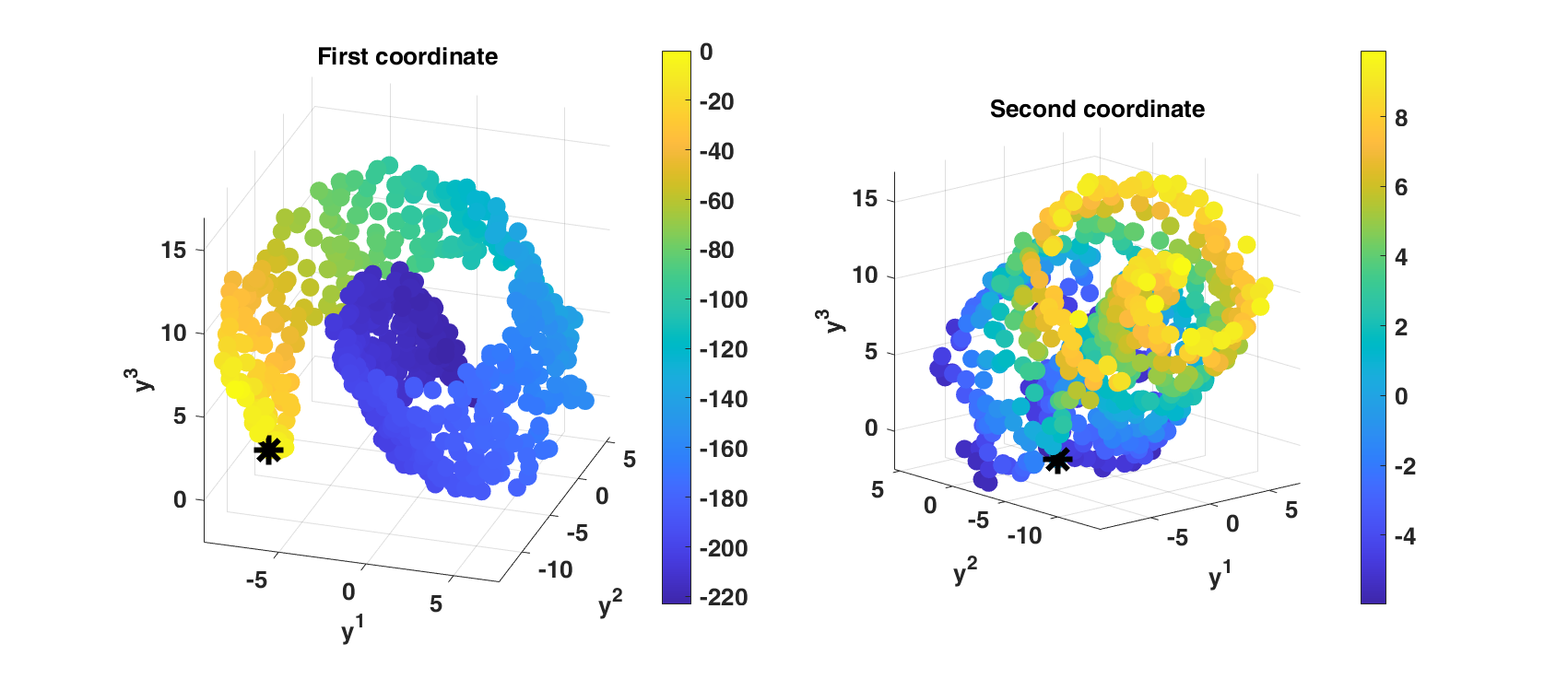}
    \caption{Right-hand side: 800 randomly chosen points lay on a Swiss Roll. The black star denotes the reference point where the latent coordinates vanish. Noise is introduced in each coordinate by adding uniformly randomly samples from the interval $[0, 0.35]$. Matrices $\mathbf{D}$ are constructed on the basis of the local tangent unit vectors.}
    \label{Fig2DatlasSR}
\end{figure}

The result of the algorithm is a global mapping $\{\bar{\mathbf{x}}_1, \bar{\mathbf{x}}_2, \dots, \bar{\mathbf{x}}_M\}$, where each global coordinate $\bar{\mathbf{x}}_i \in \mathbb{R}^d$ provides a consistent representation of the corresponding high-dimensional point in the latent space. 

Such global mapping attempts a merge of all local mappings defined by the projection matrices $\mathbf{D}_i$ into a consistent global mapping, preserving the local geometry of the latent space.

Again, it is worth stressing that the above construction of the global coordinates can be convenient in the ERRM as it allows us to take advantage of the readily available projection matrices.
While, other well-established methods can be used for the scope as well (e.g. diffusion maps \cite{coifman2005geometric}, isomap \cite{isomap2002}), an exhaustive discussion on the several possible options is beyond the scope of this work and won't be addressed here.

\subsubsection{A few illustrative examples}
The aforementioned procedure of composing the several local mapping defined by the matrices $\mathbf{D}$ into a unique consistent global mapping is tested below on a few benchmarks cases with one- and two-dimensional intrinsic dimension.
Benchmarks consists of point clouds on: i) a noisy two-dimensional circle; ii) noisy two-dimensional Archimedes spiral:
\begin{equation}
\begin{split}
y^1(\vartheta) &= (1 + 0.5\vartheta) \cos(\vartheta) \\
y^2(\vartheta) &= (1 + 0.5\vartheta) \sin(\vartheta)
\end{split}
\end{equation}
iii) a noisy three-dimensional Fermat spiral:
\begin{equation}
\begin{split}
y^1(\vartheta) &= 0.5 \sqrt{\vartheta} \cos(\vartheta) \\
y^2(\vartheta) &= 0.5 \sqrt{\vartheta} \sin(\vartheta) \\
y^3(\vartheta) &= \frac{\vartheta}{\vartheta_{\max}}
\end{split}
\end{equation}
iv) and a noisy inclined Swiss roll, defined by the equations:
\begin{equation}
\begin{split}
\tilde{y}^1(\vartheta) &= \vartheta \cos(\vartheta) \\
\tilde{y}^2(\vartheta) &= \vartheta \sin(\vartheta) \\
\end{split}
\end{equation}
with
\begin{equation}
    \vartheta = \left[ \frac{4 \pi}{2} \cdot 0.1, \frac{4 \pi}{2} \cdot 1.6 \right]; \tilde{y}^3= \left[0, 15\right]
\end{equation}
The final coordinates $y^i$ of the inclined Swiss roll are obtained from $\tilde{y^i}$ upon a rotation of an angle of $\frac{\pi}{4}$ around the axis $\tilde{y}^1$.
%
Results are reported in Figures \ref{Fig1DatlasCirle}, \ref{Fig1DatlasFermat} and \ref{Fig2DatlasSR}, where it can be seen that the constructed global mapping can correctly recover the natural coordinates of the considered manifolds.

\subsection{Alignment of the projection matrices}\label{alignDsec}
In order to ensure consistency among nearby local mappings, the vectors spanning the rows of the projection matrices $\textbf{D}$ have to be optimally aligned.
To this end, let the two matrices $\mathbf{D}_1 \in \mathbb{R}^{n \times m}$ and $\mathbf{D}_2 \in \mathbb{R}^{n \times m}$ be defined.
We compute an optimal alignment of the two matrices by using singular value decomposition (SVD).

To align the bases, we first compute the matrix $\mathbf{M} = \mathbf{D}_1^\top \mathbf{D}_2$, which captures the pairwise relationships between the basis vectors of $\mathbf{D}_1$ and $\mathbf{D}_2$.
The SVD of $\mathbf{M} = \mathbf{U} \mathbf{\Sigma} \mathbf{V}^\top$ provides orthogonal matrices $\mathbf{U}$ and $\mathbf{V}$ that can be used to calculate the optimal rotation matrix $\mathbf{R} = \mathbf{U} \mathbf{V}^\top$. 
The latter rotation matrix minimizes the Frobenius norm of the difference between the aligned basis $\mathbf{D}_2^\text{aligned}$ and the original basis $\mathbf{D}_1$, achieving:
\[
\mathbf{D}_2^\text{aligned} = \mathbf{R} \mathbf{D}_2.
\]

The result is a transformed basis $\mathbf{D}_2^\text{aligned}$ that is optimally aligned to $\mathbf{D}_1$ in the sense of minimizing the discrepancy between the two bases while preserving their geometric properties. 
%

\subsection{Active learning orchestration}\label{ALprocedure}
One possible strategy for navigating the latent space towards fixed points can be based on Gaussian process regression and properly selected acquisition and objective functions.
Clearly, this is only one possible approach among several options that will not be discussed in this work. 

Let's consider an array $\mathbf{X}$ containing $n$ observed points in latent space and an array $\mathbf{Y}$ with the corresponding function values (objective function values).
Let's define a maximum allowable step size $dx_{i,max}$ for corrections along the generic latent dimension $i$.

We aim at estimating the optimal correction $dx_{i,opt}$ for the generic dimension $i$, as well as the observed point to which the correction is to be applied.

First, a Gaussian process regression model is trained using the input data $\mathbf{X}$ and $\mathbf{Y}$.
In general, another regressor type can also be used (e.g. random forest \cite{ling2017high,trezza2022minimal}) nonetheless, for the sake of simplicity, in this work we only focus on Gaussian processes.
Here, we have used a regression model employing a squared exponential kernel with a properly chosen kernel parameter $\texttt{kernelScale}$. 
Those models allow the prediction of a mean value $\mu$ and standard deviation $\sigma$ for the objective function at any given point.



To identify the optimal step adjustments, we evaluate a grid of candidate adjustments $dx_i$ each within the range $[-dx_{max,i}, dx_{max,i}]$. 
For each observed point in $\mathbf{X}$, the algorithm iterates over this grid of candidate adjustments.
For each possible correction vector $[dx_{1},...,dx_{d}]$, the candidate point is calculated as
\begin{equation}
    \mathbf{x}_{new} = \mathbf{x} + [dx_{1},...,dx_{d}],
\end{equation}

At each of those new points, the Gaussian process model (or possibly other regression models) can be used to predict the mean $\mu$ and standard deviation $\sigma$ of the objective function.
These predictions are then used to compute the value of the acquisition function, which quantifies the utility of the candidate point for exploration or exploitation. 
In the following, we only review some of the most popular choices for the acquisition function (without pretending to be exhaustive):

\textbf{Maximum Likelihood of Improvement (MLI)}: This function evaluates the likelihood of achieving an improvement over the threshold (i.e. the current maximum value in $\mathbf{Y}$) and is given by
\begin{equation}
\texttt{acquisition} = \Phi\left(\frac{\mu - \texttt{threshold}}{\sigma}\right),
\end{equation}
where $\Phi$ is the cumulative distribution function of the standard normal distribution.

\textbf{Expected Improvement (EI)}: This function evaluates the expected value of improvement and is given by:
\begin{equation}
    \texttt{acquisition} = (\mu - \texttt{threshold}) \Phi\left(\frac{\mu - \texttt{threshold}}{\sigma}\right) + \sigma \phi\left(\frac{\mu - \texttt{threshold}}{\sigma}\right),
\end{equation}

where $\phi$ is the probability density function of the standard normal distribution.
%
%

\textbf{Upper Confidence Bound (UCB)}: This function balances exploration and exploitation and is given by
\begin{equation}
\texttt{acquisition} = \mu + \kappa \sigma,
\end{equation}

with $\kappa$ being a tunable parameter controlling the trade-off between exploitation and exploration.

For each candidate point, the acquisition function value is compared with the current best acquisition value.
When the new acquisition function value is greater, the best values are updated to reflect the new candidate point and its corresponding adjustments. 
This process continues until all candidate adjustments for all points in $\mathbf{X}$ have been evaluated.

As a result, we identify the optimal step adjustments $[dx_{best,1},...,dx_{best,d}]$ along with the corresponding point in $X$ to which this correction has to be applied. 
The latter corrected point is the new point in the latent space to which the procedure jumps, as pictorially illustrated in Figure (\ref{fig2}) (d).

\section{RESULTS}\label{results}
\subsection{A first test case in 2D: The M\"{u}ller-Brown example}
As a first example to test the above explorative RRM, we consider the following very popular M\"{u}ller-Brown potential energy:
\begin{equation}\label{MBpotential}
V(y^1, y^2) = \sum_{i=1}^{4} A_i \exp \left[ a_i (y^1 - y^1_{0i})^2 + b_i (y^1 - y^1_{0i})(y^2 - y^2_{0i}) + c_i (y^2 - y^2_{0i})^2 \right]
\end{equation}
with the parameters defined in Table \ref{tabparam}.
\begin{table}[h!]
\centering
\begin{tabular}{|c|c|c|c|c|c|c|}
\hline
$i$ & $A_i$ & $a_i$ & $b_i$ & $c_i$ & $y^1_{0i}$ & $y^2_{0i}$ \\
\hline
1 & -200 & -1.0 & 0 & -10.0 & 1.0 & 0.0 \\
2 & -100 & -1.0 & 0 & -10.0 & 0.0 & 0.5 \\
3 & -170 & -6.5 & 11 & -6.5 & -0.5 & 1.5 \\
4 & 15 & 0.7 & 0.6 & 0.7 & -1.0 & 1.0 \\
\hline
\end{tabular}
\caption{Parameters of the Müller–Brown potential. Note that superscripts do not denote power. \label{tabparam}}
\end{table}
Furthermore, we assume that the dynamical system (\ref{sistemadinamicogenerale}) takes the more explicit form:
\begin{equation}\label{MBode}
\left\{
    \begin{aligned}
     \frac{dy^1}{dt} &= -\frac{\partial V}{\partial y^1} (y^1,y^2), \\
     \frac{dy^2}{dt} &= -\frac{\partial V}{\partial y^2} (y^1,y^2).
\end{aligned}
\right.
\end{equation}

\subsubsection{One-dimensional SIM}
\begin{figure}[htbp]
    \centering
    \begin{subfigure}{0.48\textwidth}
        \centering
        \includegraphics[width=\textwidth]{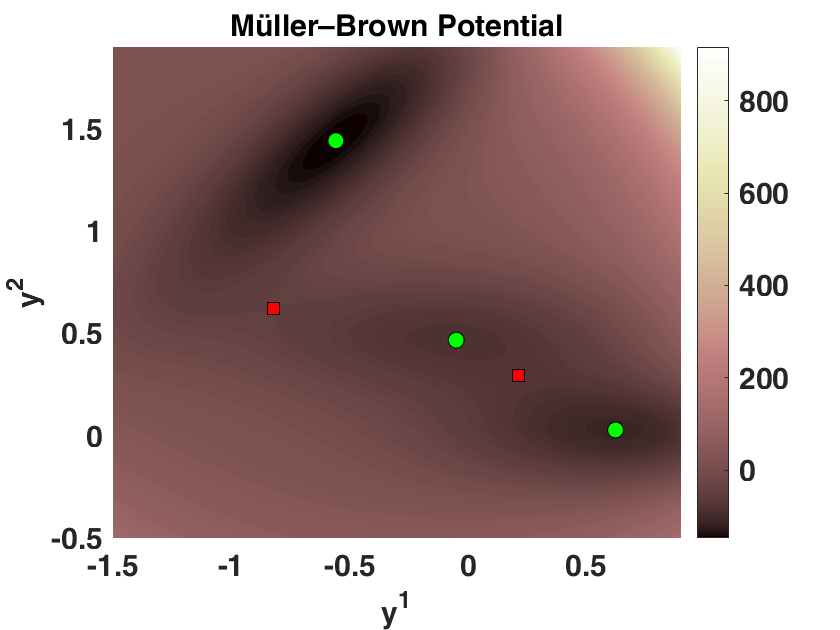}
    \end{subfigure}
    \begin{subfigure}{0.48\textwidth}
        \centering
        \includegraphics[width=\textwidth]{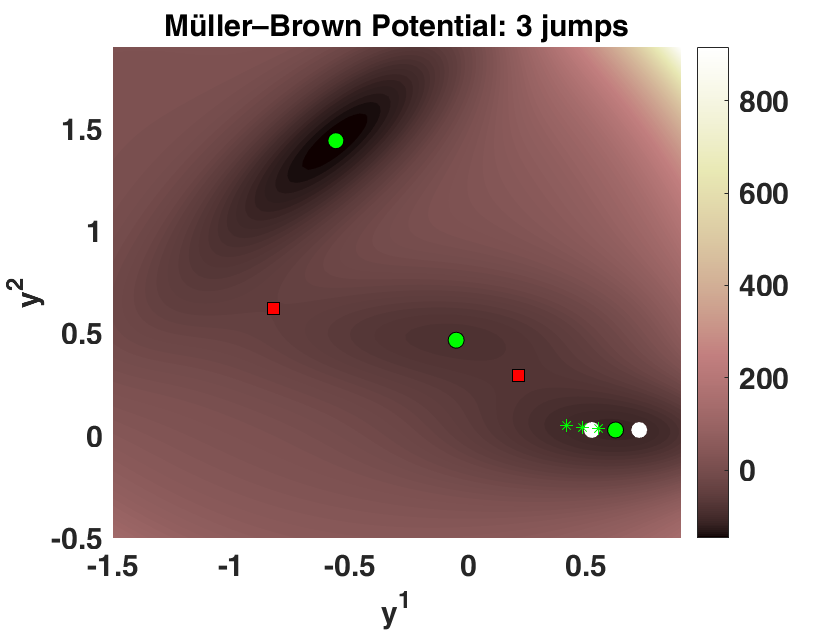}
    \end{subfigure}
    
    \vspace{0.5cm}
    
    \begin{subfigure}{0.48\textwidth}
        \centering
        \includegraphics[width=\textwidth]{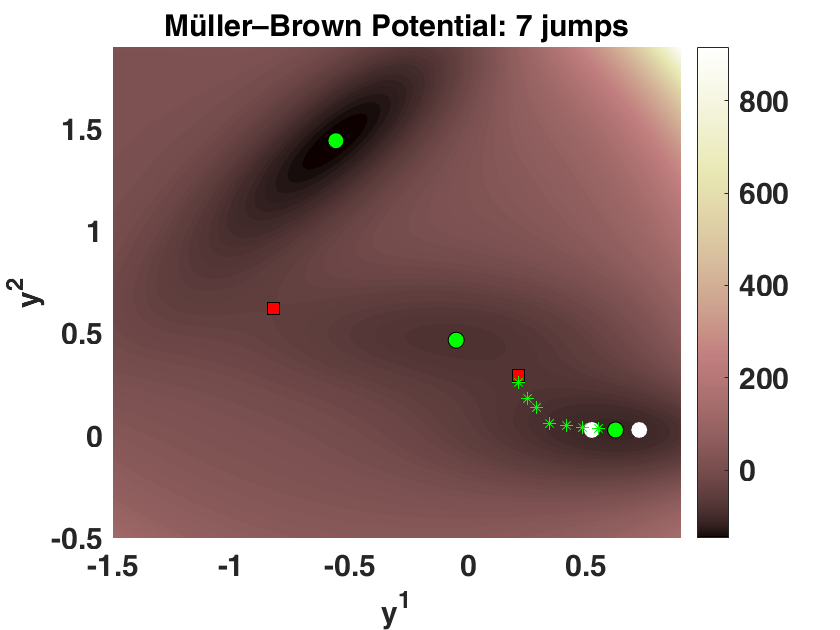}
    \end{subfigure}
    \begin{subfigure}{0.48\textwidth}
        \centering
        \includegraphics[width=\textwidth]{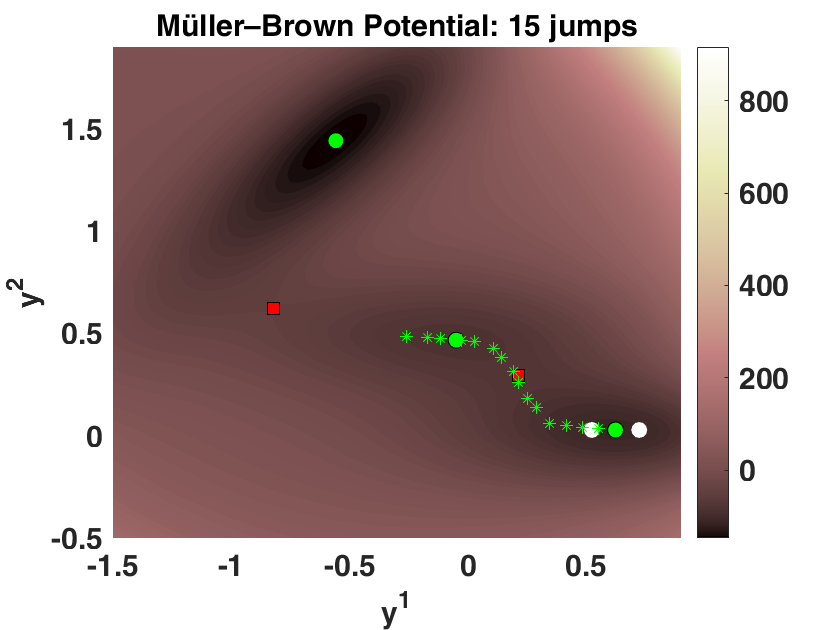}
    \end{subfigure}
    
    \vspace{0.5cm}
    
    \begin{subfigure}{0.48\textwidth}
        \centering
        \includegraphics[width=\textwidth]{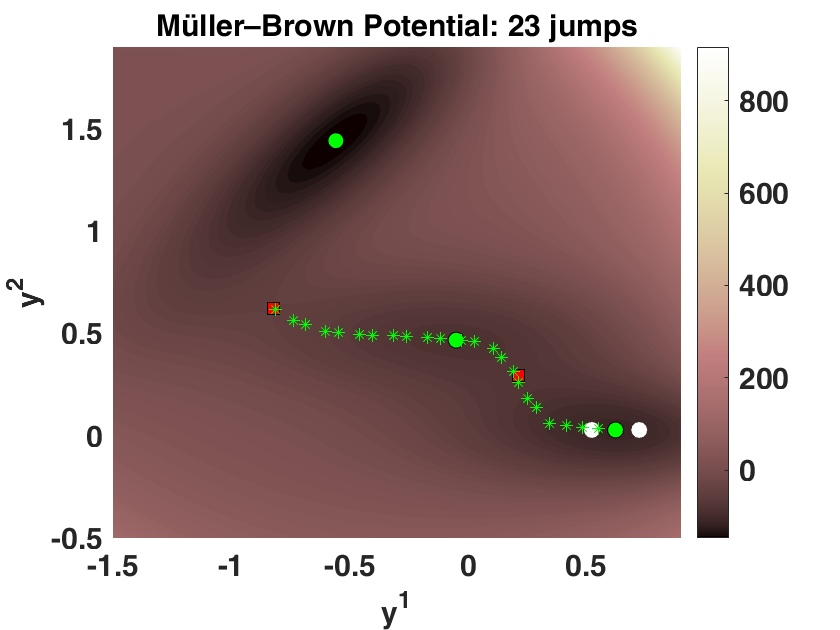}
    \end{subfigure}
    \begin{subfigure}{0.48\textwidth}
        \centering
        \includegraphics[width=\textwidth]{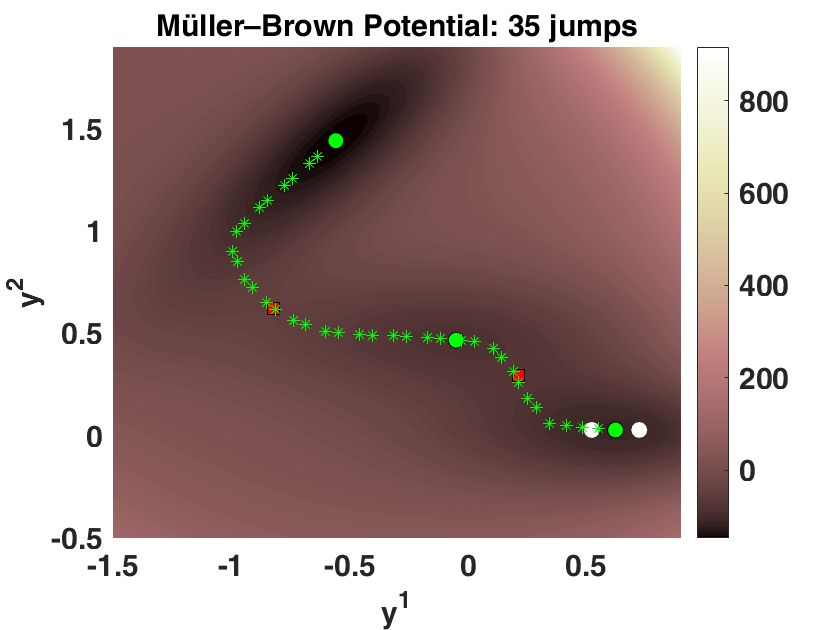}
    \end{subfigure}
    
    \caption{Gradual discovery of fixed points for system (\ref{MBode}). (a) Isocontours of the potential with fixed points. (b) The ERRM is initiated from the bottom right fixed point (assumed to be known), and the white circles represent the initial guess of the two secondary ($N_s\ge2$) points associated with the starting configuration. First three jumps are shown. (c) After 7 jumps the first saddle point is reached. (d) After 15 jumps the procedure is beyond the second stable point. (e) After 23 jumps the procedure has reached the second saddle point. (f) After 35 jumps the procedure has reached the last fixed point. Green stars represent pivots only. Except for the initial point, secondary points are omitted.}
    \label{fig:matrice3x2}
\end{figure}

In Figure \ref{fig:matrice3x2}, we show a possible instance of the explorative RRM approach on the popular M\"{u}ller-Brown example where the dimension of the SIM has been set to unity ($d=1$).
The specific choice of parameters from Table \ref{tabparam} ensures three stable fixed points (green circles) and two saddle points (red squares).
%
%
For the specific example in Figure \ref{fig:matrice3x2}, the RRM procedure is implemented by using the Dormand-Prince method during the relaxation step for a time $\tau=0.5 \times 10^{-6}$ before implementing the redistribution step (for both the slow and fast subspace relaxation). 
The RRM is terminated when the movement of the pivot ($\Delta \mathbf{c}_p$) between two subsequent RRM steps is sufficiently small as compared to the relaxation movement only.
More specifically, we impose: $\lvert \Delta \mathbf{c}_p \rvert / \lvert \mathbf{f} \tau \rvert < 10^{-3}$.
Finally, the explorative step (see Figure \ref{fig2} (d)) is achieved by shifting the projection of the current pivot and secondary points ($x_i$) in the latent space by a fixed quantity: $x_i \pm \delta x_0$, with $\delta x_0=0.07$.
%
\begin{figure}[htbp]
    \centering
    \begin{subfigure}{0.48\textwidth}
        \centering
        \includegraphics[width=\textwidth]{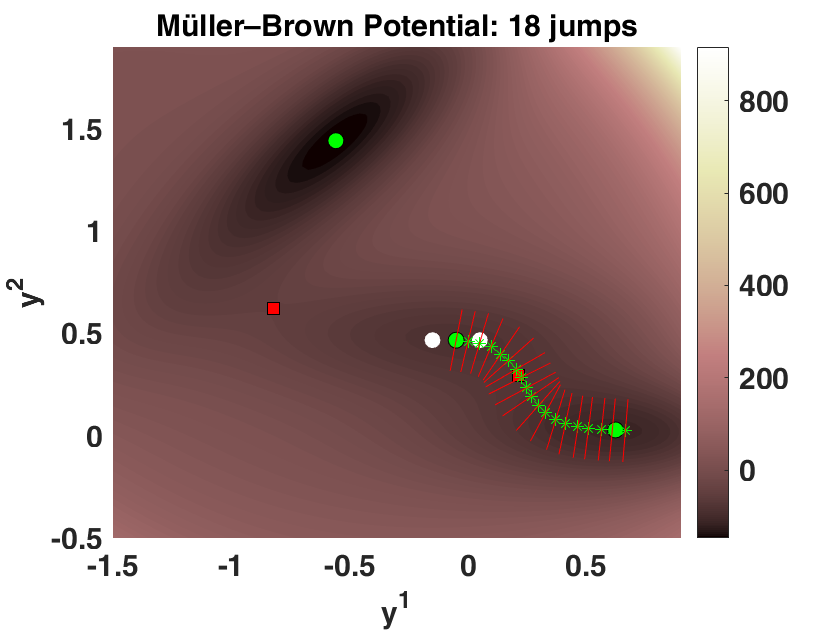}
    \end{subfigure}
    \begin{subfigure}{0.48\textwidth}
        \centering
        \includegraphics[width=\textwidth]{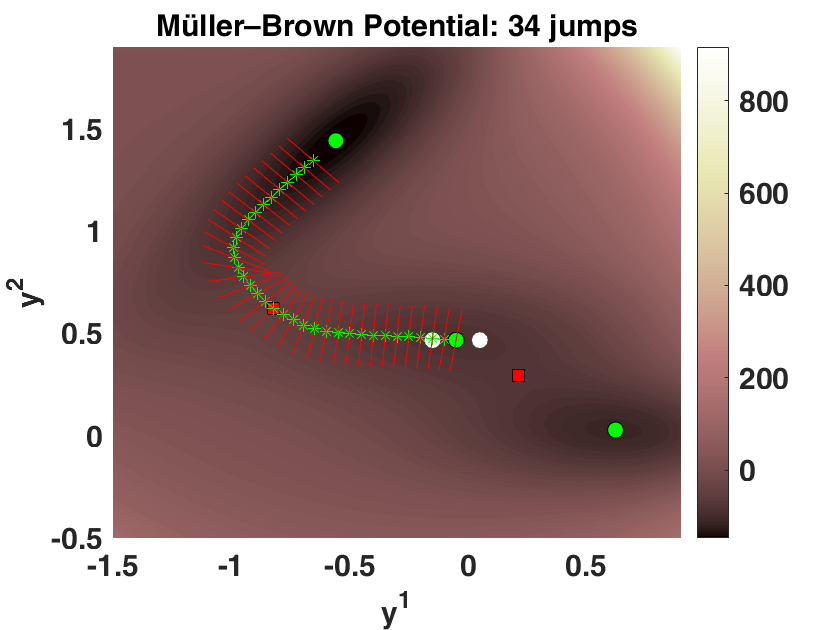}
    \end{subfigure}
    
    \vspace{0.5cm}
    
    \begin{subfigure}{0.48\textwidth}
        \centering
        \includegraphics[width=\textwidth]{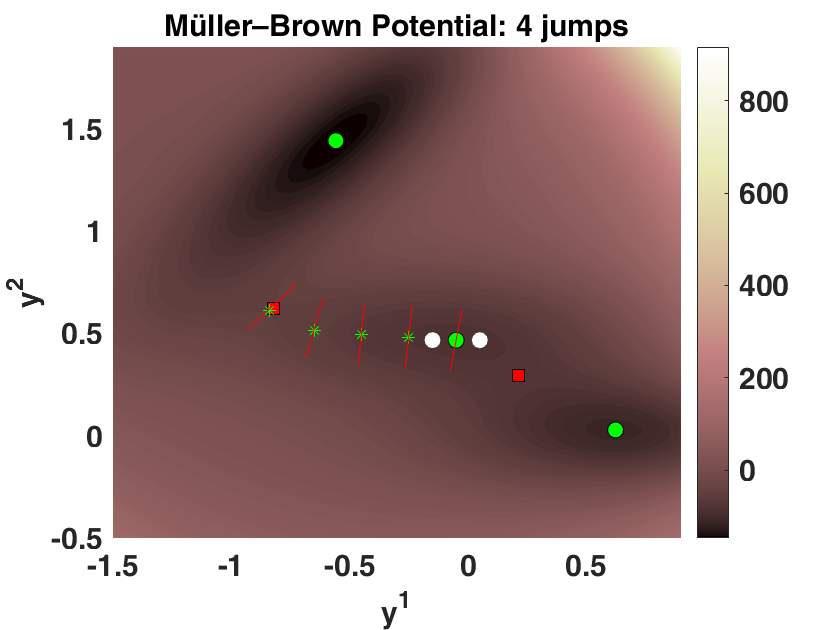}
    \end{subfigure}
    \begin{subfigure}{0.48\textwidth}
        \centering
        \includegraphics[width=\textwidth]{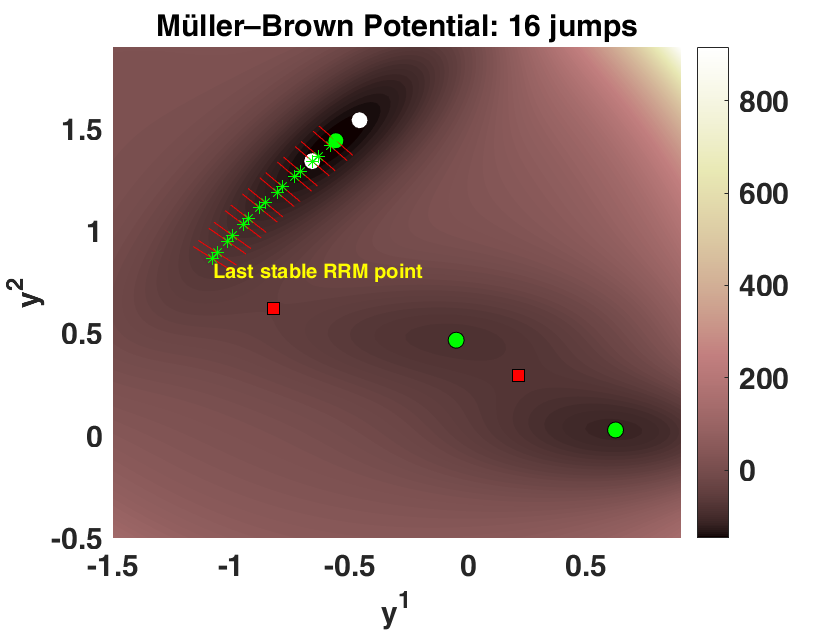}
    \end{subfigure}
    
    \caption{Gradual discovery of the fixed points in the case of a M\"{u}ller-Brown potential energy. (a) The middle stable fixed point is chosen as starting point with $\delta x_0=0.05$ and the right-most unstable and stable fixed points discovered. (b) The middle stable fixed point is chosen as starting point with $\delta x_0=-0.05$ and the unstable and stable fixed points are discovered in the upper part of the phase space. (c) The efficiency of the method (i.e. the necessary steps to reach new configurations) ca be improved by increasing with larger jumps. Here we choose $\delta x_0=-0.2$ and only four jumps are sufficient to reach the left-most saddle point. (d) While the left-most saddle point can be reached by starting from both the right-most and the second right-most stable fixed point, this does not occur when starting from the left-most stable point. In the figure we stop at the last point where the RRM converges successfully. In all sub-figures, the red segments represent the RRM refined fast subspace at all reported pivots (green stars).}
    \label{fig:matrice2x2}
\end{figure}
In Figure \ref{fig:matrice2x2}, we apply the explorative RRM method with different starting points and different values of the jump parameter $\delta x_0$.  

\subsubsection{Higher-dimensional SIM}\label{2DMBsec}
As visible in Figure \ref{fig:matrice2x2} (d), when imposing $d=1$, there is no guarantee that the explorative RRM can reach a saddle point starting from a minimum of the energy potential (i.e. a fixed stable point).
In fact, it may happen that the one-dimensional RRM method does not converge. 
In the latter case, we suggest that the method illustrated in Figure \ref{fig2} is to be implemented in a higher dimensional reduced space (i.e. $d>1$), namely the intrinsic dimension of the energy function support.
As suggested in \cite{chiavazzo2012approximation}, the lowest value of $d$ that ensures convergence of the RRM method reveals the dimension of the SIM.
Clearly, in the two-dimensional test case (\ref{MBode}) we have $d=2$ and according to (\ref{Nspoints}) $Ns \ge 5$.
Furthermore, since $d=n$, in the considered test case we do not need to \textit{surf} on a SIM. 
In other words, we don't need to implement the procedure in Figure (\ref{fig2}), and we only focus on the navigation strategy toward unknown fixed points.
To this end, we suggest that the search for fixed points proceeds following an active learning orchestrated procedure that attempts the optimization of a properly selected objective function. 
That strategy can start from the last point where the RRM was successful, or even from the initially available fixed stable point.
%
\begin{figure}[htbp]
    \centering
    \begin{subfigure}{0.48\textwidth}
        \centering
        \includegraphics[width=\textwidth]{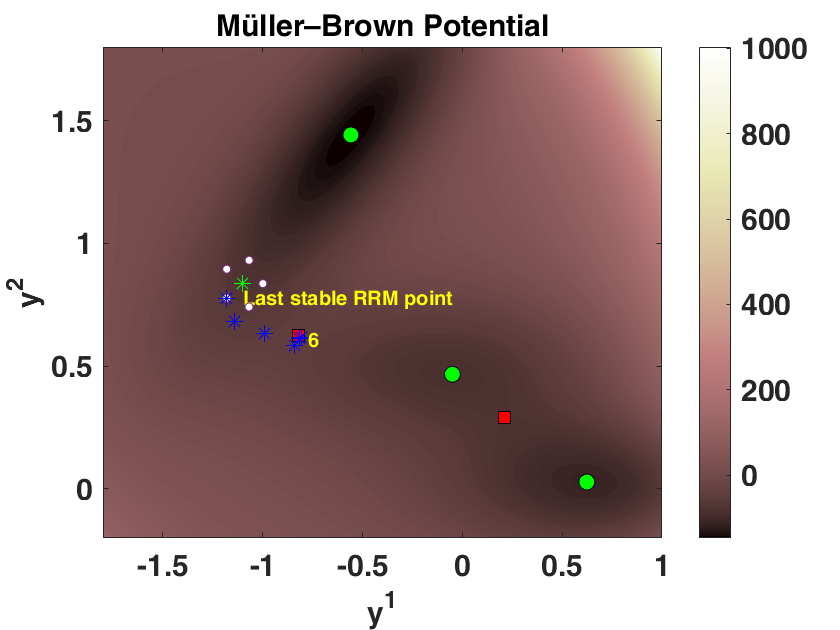}
    \end{subfigure}
    \begin{subfigure}{0.48\textwidth}
        \centering
        \includegraphics[width=\textwidth]{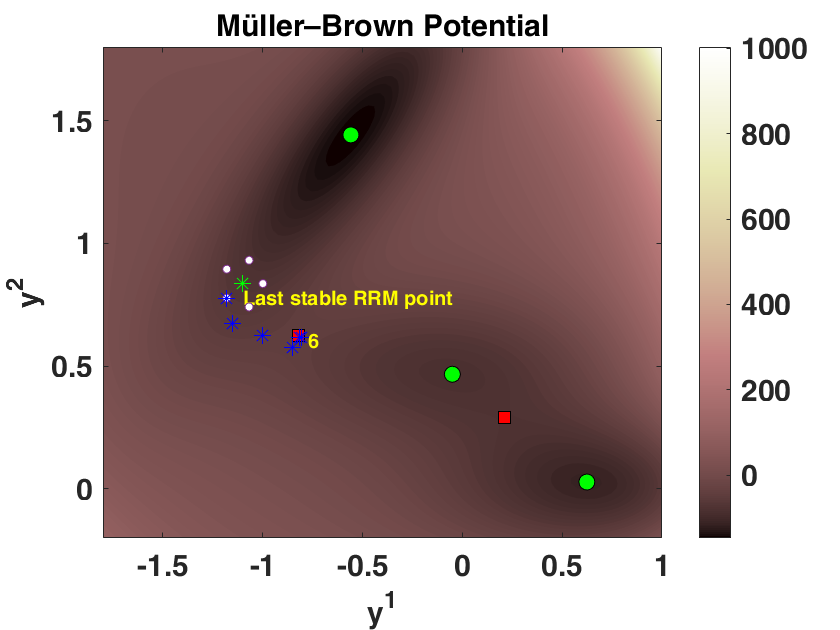}
    \end{subfigure}
    
    \vspace{0.5cm}
    
    \begin{subfigure}{0.48\textwidth}
        \centering
        \includegraphics[width=\textwidth]{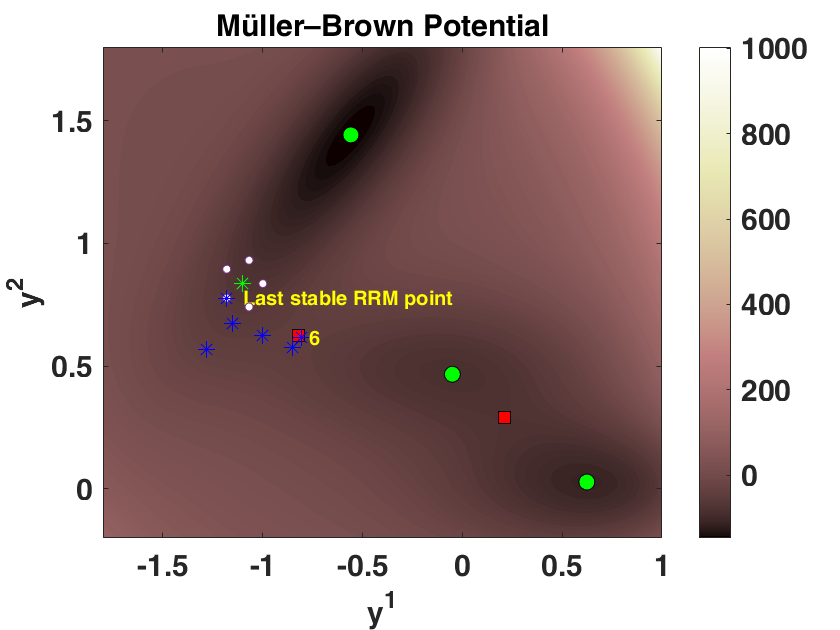}
    \end{subfigure}
    \begin{subfigure}{0.48\textwidth}
        \centering
        \includegraphics[width=\textwidth]{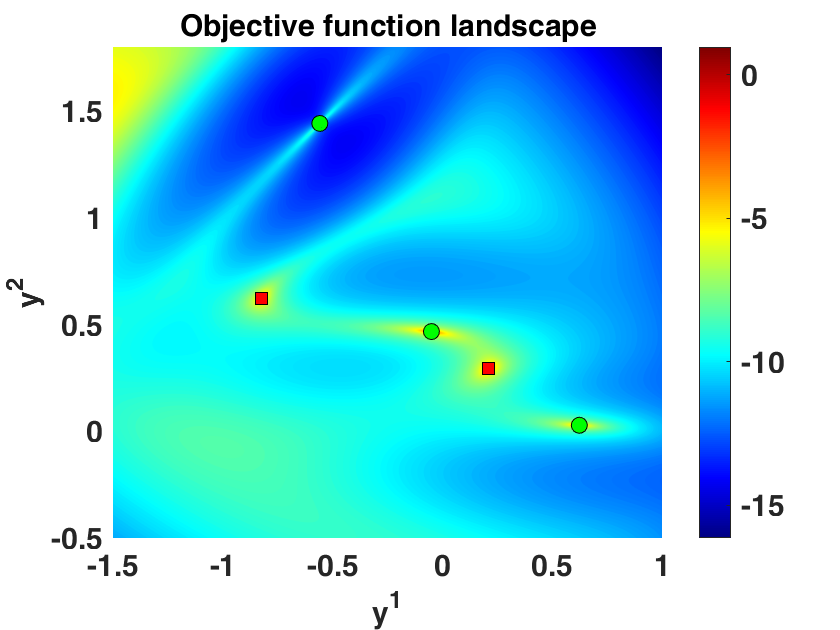}
    \end{subfigure}
    
    \caption{(a-c) Starting from the last point of the RRM procedure in Figure \ref{fig:matrice2x2} (d), we implement an active learning search of a saddle point. In six steps the saddle point is reached using the Upper Confidence Bound as an acquisition function ($\kappa=1.5-5$). (d) The objective function landscape. White circles are the secondary points at the starting point, while blue stars represent pivots only. For the sake of clarity, all other secondary points are not represented.}
    \label{fig:AL:UCB}
\end{figure}

%
\begin{figure}[htbp]
    \centering
    \begin{subfigure}{0.48\textwidth}
        \centering
        \includegraphics[width=\textwidth]{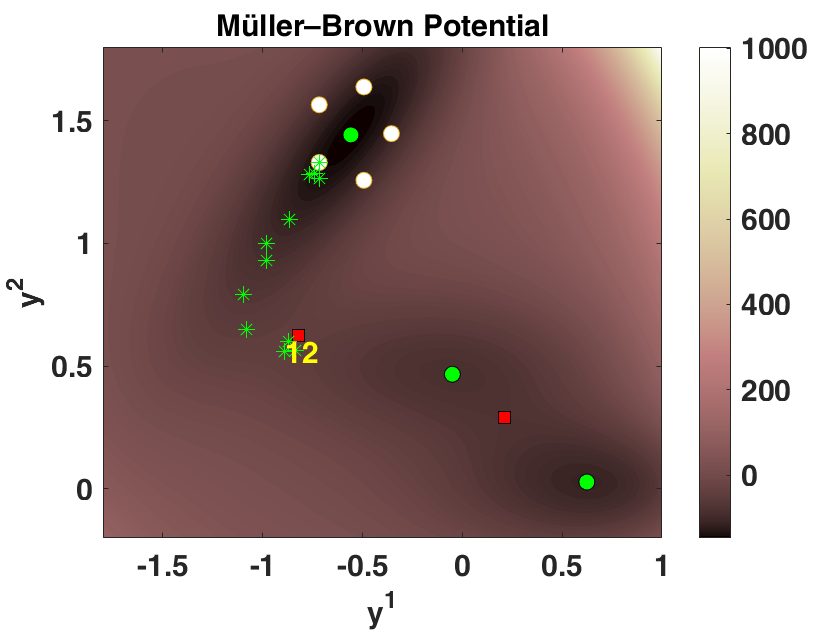}
    \end{subfigure}
    \begin{subfigure}{0.48\textwidth}
        \centering
        \includegraphics[width=\textwidth]{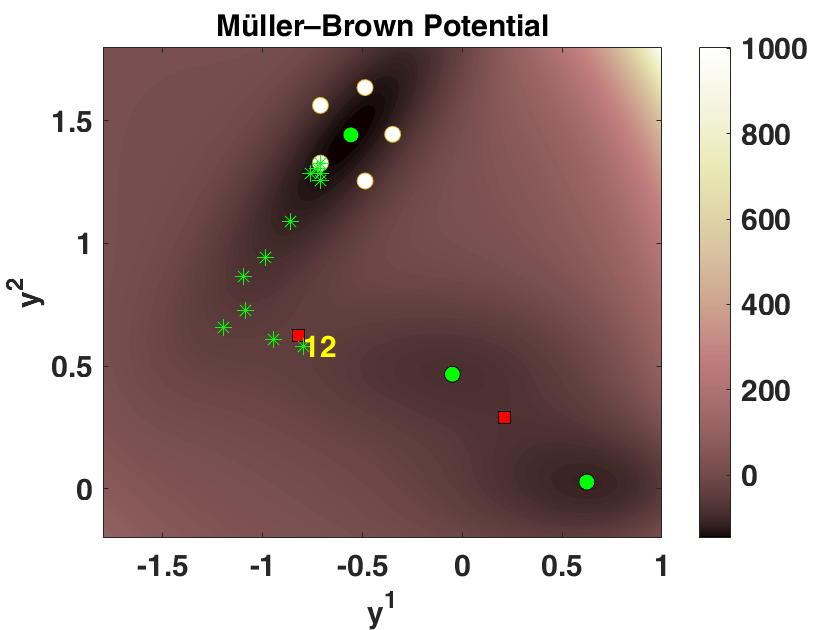}
    \end{subfigure}
    
    \vspace{0.5cm}
    
    \begin{subfigure}{0.48\textwidth}
        \centering
        \includegraphics[width=\textwidth]{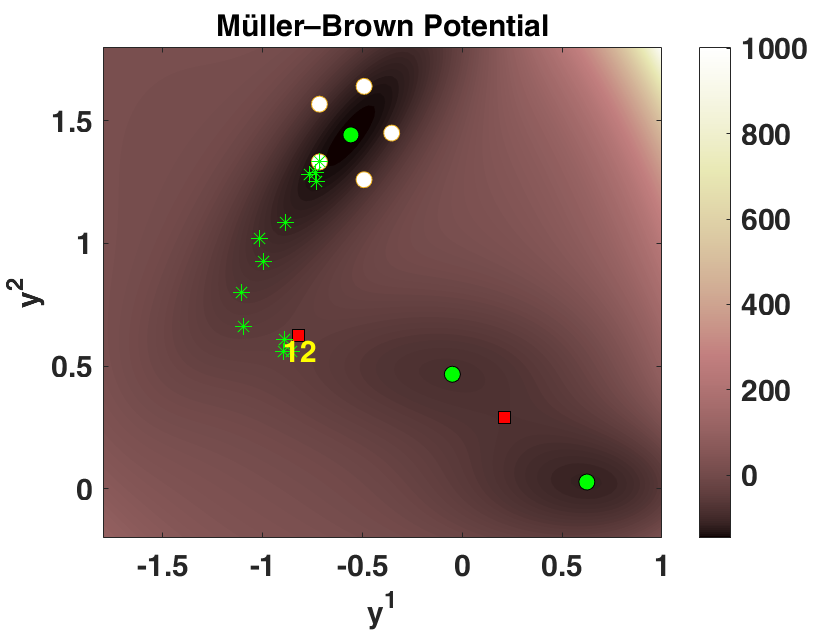}
    \end{subfigure}
    \begin{subfigure}{0.48\textwidth}
        \centering
        \includegraphics[width=\textwidth]{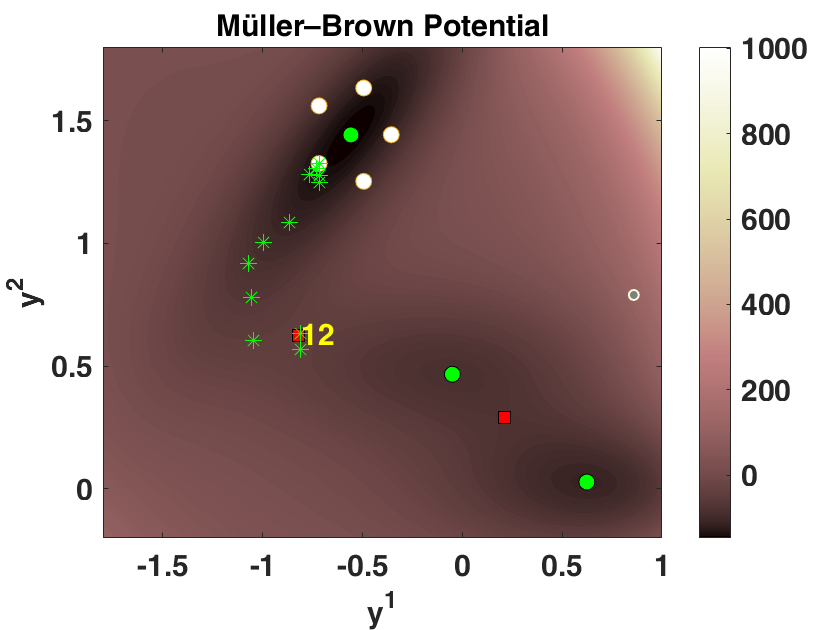}
    \end{subfigure}
    
    \caption{Four different instances of an active learning orchestrated navigation from the upper most minima towards the middle saddle point. The Expected Improvement (EI) is the activation function used. White circles are the secondary points at the starting minimum. For the sake of clarity, all other secondary points are not represented.}
    \label{fig:AL:EI}
\end{figure}
We follow the active learning navigation described in Section \ref{ALprocedure} with the following objective function to maximize: 
\begin{equation}\label{objfun}
    \text{Obj} = \ln\left(\frac{d_l^\alpha}{(\|\nabla V\| + \epsilon)^\beta}\right)
\end{equation}
where $d_l$ and $\nabla V$ represent the distance in the latent space and the gradient of the energy function, respectively.
Furthermore, the model parameters $\alpha$ and $\beta$ are to be properly chosen, whereas $\epsilon$ is a small positive quantity that can be adopted for avoiding possible division by zero.
The main rationale behind Eq. (\ref{objfun}) is that - starting from the already known stable configuration - the active learning orchestration is expected to steer the navigation towards regions with a vanishing gradient of the energy function that are also sufficiently distant from the starting point. 
Clearly, there is no unique way of defining such an objective function, and most importantly there is no guarantee that the function (\ref{objfun}) is optimal and of any general relevance.
More effective options might be suggested in future works.

As visible in Figure \ref{fig:AL:UCB}, starting from the last point where the one-dimensional RRM is stable, the active learning orchestrated search successfully reaches the saddle point after six steps. 
For this case, we selected: $\alpha=1$, $\beta=2$, $dx_{i,max}=0.01$.
For the Gaussian regressor, we use an adaptive kernel parameter where,  $\text{kernelScale}=0.1$ unless when between two subsequent steps the distance between the initial pivot $\textbf{c}_0$ and the corrected one $\textbf{c}_1$ is smaller than a threshold ($\| \textbf{c}_0-\textbf{c}_1 \| \le 0.05$). 
In the latter case, to favor visiting unknown regions, we double the current kernel parameter until two subsequent steps deliver $\| \textbf{c}_0-\textbf{c}_1 \|>0.05$ so that the initial $\text{kernelScale}=0.1$ is restored.
In Figure \ref{fig:AL:EI}, we show four different instances of a navigation process starting directly from upper-most minimum of the energy potential.
In this case, twelve steps are sufficient to reach the saddle point.
In all cases reported in Figures \ref{fig:AL:UCB} and \ref{fig:AL:EI}, the secondary points are generated by selecting five uniformly distributed samples on a circle of radius $r=0.12$ around the starting pivot.

\subsection{A double-well test case on a cylinder in 3D}
As a second higher-dimensional test case, we consider the system presented in Ref. \cite{georgiou2017exploration}, where a two-dimensional SIM (i.e. a cylinder of radius $R$ embedded in three-dimensional phase space) acts as a support for a two-well energy function. In particular, we consider the following stochastic differential equations (SDEs):
\begin{equation}\label{SDE01}
    \begin{aligned}
    \frac{dy^1}{dt} &= -\frac{1}{\epsilon} \left( y^1 - R \cos\left(\frac{\pi}{2} - \theta\right) \right) + \cos(\theta) \left(-4cR\theta (R\theta - 1)(R\theta + 1) - by^2 \right) + D\sqrt{2} dW_1, \\
    \frac{dy^2}{dt} &= -2ay^2 - bR\theta + D\sqrt{2} dW_2,  \\
    \frac{dy^3}{dt} &= -\frac{1}{\epsilon} \left( y^3 - R \sin\left(\frac{\pi}{2} - \theta\right) \right) - \sin(\theta) \left(-4cR\theta (R\theta - 1)(R\theta + 1) - by^2 \right) + + D\sqrt{2} dW_3,
\end{aligned}
\end{equation}
where
\begin{equation}
 \theta =
\begin{cases}
    \arctan\left(\frac{y^1}{y^3}\right), & \text{if } y^1 > 0 \text{ and } y^3 > 0, \\
    -\frac{\pi}{2} - \arctan\left(\frac{y^3}{y^1}\right), & \text{if } y^1 < 0 \text{ and } y^3 < 0, \\
    \frac{\pi}{2} + \arctan\left(-\frac{y^3}{y^1}\right), & \text{if } y^1 > 0 \text{ and } y^3 < 0.
\end{cases}
\end{equation}

and the following parameter:
\[
\begin{array}{|c|c|}
\hline
\textbf{Parameter} & \textbf{Value} \\
\hline
a & 200 \\
b & -80 \\
c & 20 \\
R & \frac{2}{\pi} \\
\epsilon & 10^{-4} \\
\hline
\end{array}
\]
with $W1$, $W2$, $W3$ being independent standard Brownian motions.
By construction, the system admits a two-dimensional SIM with radius R and main axis $y^2$.
\begin{figure}[h]
    \centering
        \includegraphics[width=0.95\textwidth, height=2.5in]{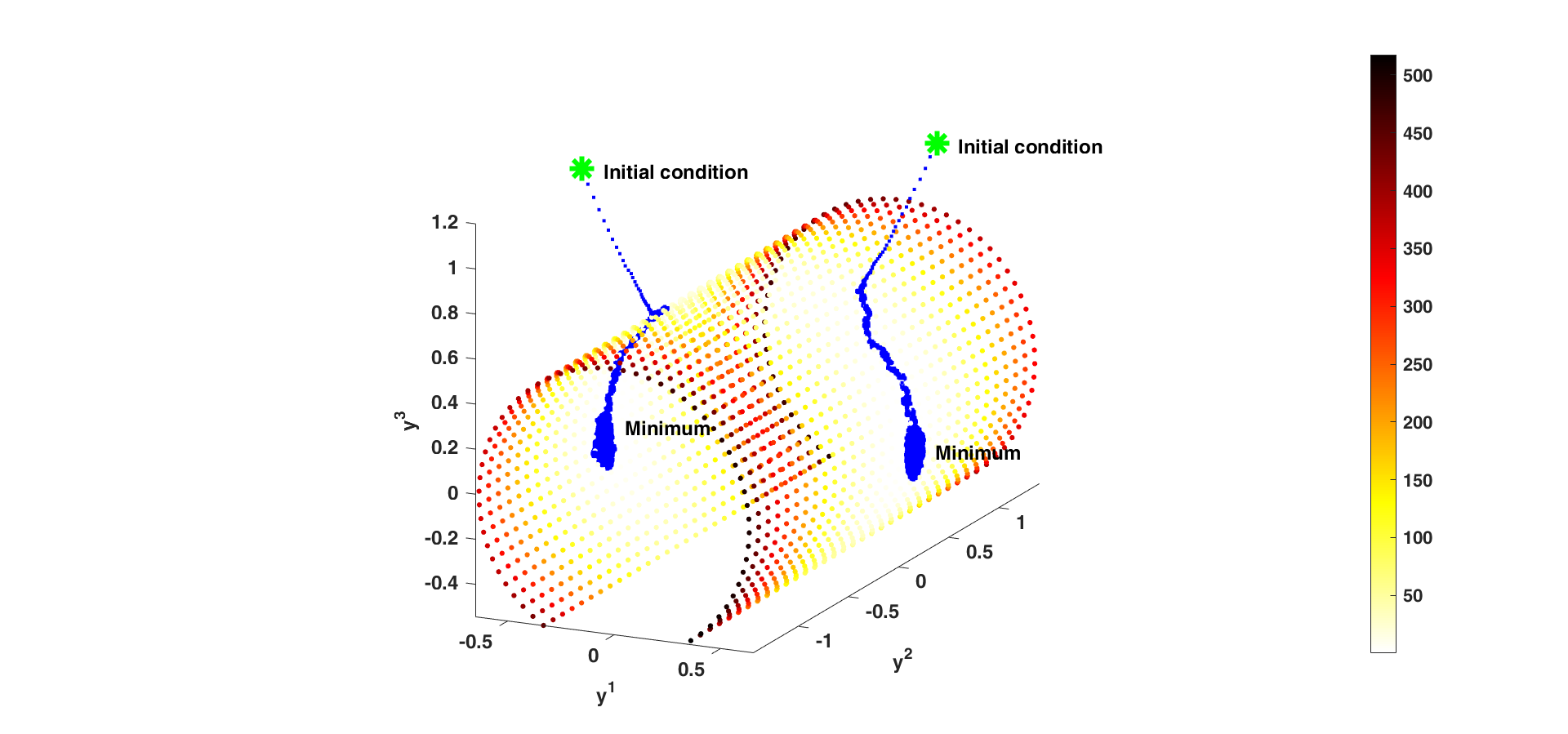}
        \caption{Two solution trajectories of the SDE (\ref{SDE01}) starting from the initial conditions: $[-0.6,-0.5,1.2]$ and $[0.6,0.5,1.2]$. The reached minimum depends on the initial condition. Here, the SDE is solved by means of an Euler–Maruyama scheme for a total time $T=1$ and $\Delta t=5 \times 10^{-6}$. The colormap on the cylinder denotes the effective energy function values according to the reported colorbar. Here we used $D=0.5$.\label{figEx3d}}
\end{figure}
As visible in Figure \ref{figEx3d}, solution trajectories that start off the SIM are quickly attracted to the SIM and subsequently proceed towards one of the  minima of the effective energy landscape.
In particular, the latter SDE system has two metastable states, with trajectories arriving at one of the metastable wells and remaining {\it trapped} in their neighborhood for periods of time that are typically much longer than those of the considered simulation.

The above system can be considered as a prototypical benchmark for testing the explorative RRM method in both the one- and two-dimensional implementation, as discussed below.

\subsubsection{One-dimensional implementation}\label{Cyl1Dimp}
\begin{figure}[h]
    \centering
        \includegraphics[width=0.98\textwidth, height=3.5in]{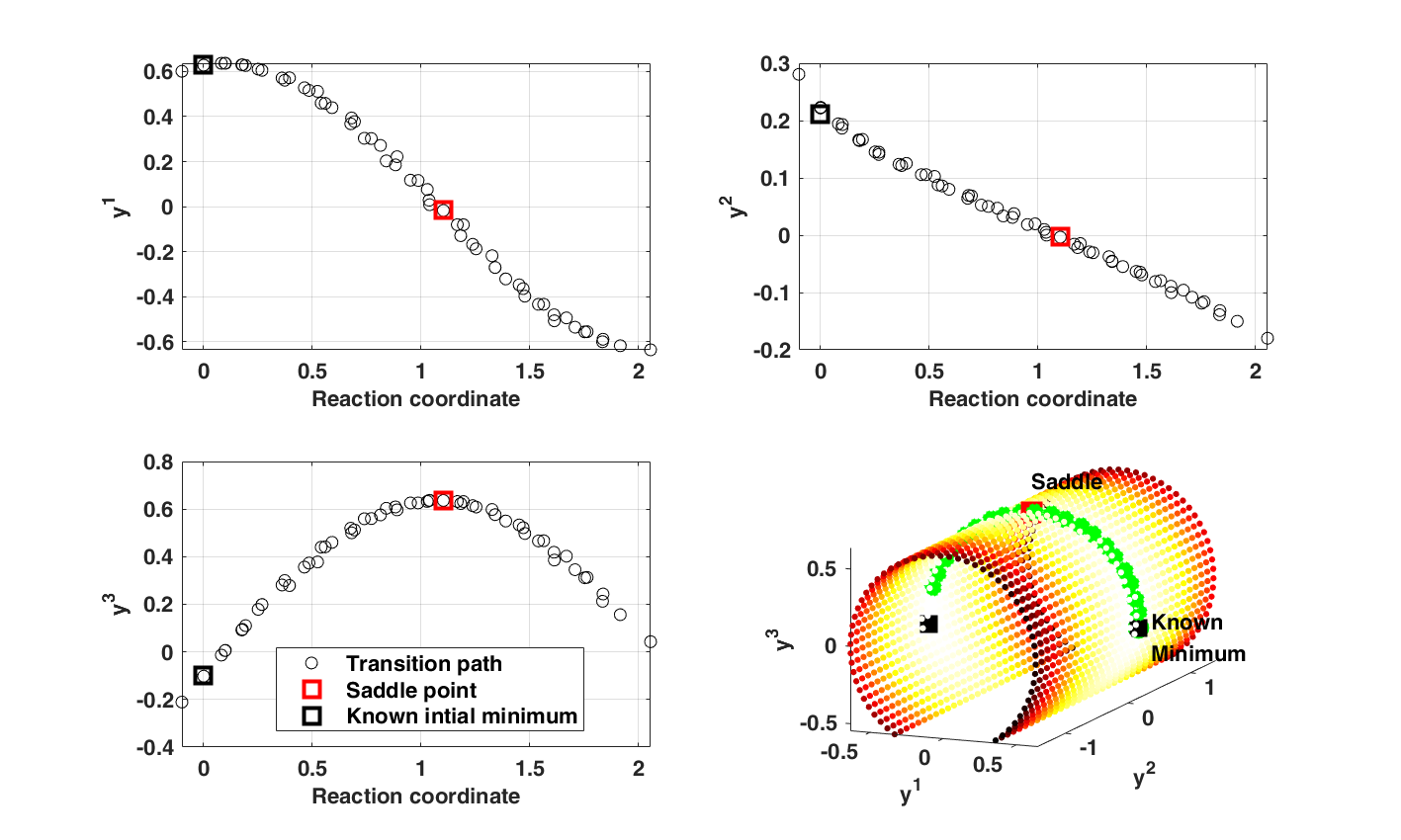}
        \caption{Explorative RRM with $d=1$ applied to the the SDE system (\ref{SDE01}). Here, we start from one of the two minima (black square, assumed as already known) and let the explorative RRM gradually reach the saddle point (red square). In the first three sub-panels, we report both pivots and secondary points. In the fourth sub-panels, for the sake of clarity, we only report the pivots (in green).\label{1dSDE}}
\end{figure}
In Figure \ref{1dSDE}, we report the implementation of the explorative RRM with $d=1$ applied to the SDE system (\ref{SDE01}).
Here, we assume that one of the two metastable minima is known in advance (i.e. the one with $y^1>0$) and the aim is to gradually discover both the saddle point and the second unknown metastable point, possibly unveiling the transition path linking one metastable point to the other.
The system (\ref{SDE01}) is solved by means of an  Euler–Maruyama scheme with $\Delta t=2 \times 10^{-5}$, hence for this case we assume not to have direct access to the gradient of the effective energy function.

In other words, we operate here with a {\it black-box} forward simulator of the SDE (\ref{SDE01}).
Hence, the relaxation step of the RRM described in the subsection \ref{slowRRMalg} above is implemented as follows: i) the pivot and all the secondary nodes are relaxed by a multiple of the time step $\Delta t$, say $q \times \Delta t$, thus generating a very short trajectory at each point; ii) As represented in Figure \ref{fig1} (b), the step i) is repeated several times (i.e. $N_r$ times); iii) for each point, the final position (to be used in the subsequent redistribution step) is the average end point over the $N_r$ short trajectories.
Similarly to the above procedure, the RRM is terminated when the distance between two subsequently refined pivot locations is small as compared to the movement of the pivot due to the relaxation only:
\begin{equation}\label{defectinvariance}
   \lvert \Delta \mathbf{c}_{RRM} \rvert / \lvert \Delta \mathbf{c}_{Rel} \rvert < 10^{-4}
\end{equation}
For the case in Figure \ref{1dSDE}, we used $D=0.5$, $N_r=20$, $q=5$. $\delta x_0=-0.1$, $\Delta t= 2 \times 10^{-5}$ and the initial parameterization matrix $\textbf{D}_{in}=[0.228,0.180,0.957]$. 
The secondary points are initially generated by a small perturbation of the pivot along $\textbf{D}_{in}$: $\textbf{c}_{sec}=\textbf{c}_{0} \pm 0.1  \textbf{D}_{in}$.
As visible in Figure \ref{1dSDE}, the ERRM is able to reach the saddle point in nearly 10 jumps.
Nonetheless, it is worth noticing that the efficiency of the method (i.e. the number of {\it jumps} needed to reach the saddle point) can be further improved by increasing the absolute value of $\delta x_0$.
More specifically, with $\delta x_0=-0.25$ here only four steps of the ERRM suffice to reach the saddle point.
In this example, the local update of the parametrization matrices $\textbf{D}$ is not accomplished by computing the null space of the estimated fast subspace, but rather considering at each refined pivot point the tangent space to the (\ref{tensoriallift}).
This is acceptable as the local low-dimensional parametrization (\ref{projectionontomanifold}) is not unique.

Finally, as reported in the first three sub-panels of Figure \ref{1dSDE}, the knowledge of all the local parametrization matrices $\textbf{D}$ allows the generation of a global mapping of the transition path from the known metastable minimum to the unknown one by using the algorithm described in Section \ref{secatlasconst}, where we impose that the global reaction coordinate vanishes at the first pivot $\textbf{c}_0=[0.63,0.21;-0.10]$.

\begin{figure}[htbp]
    \centering
    \begin{subfigure}{0.9\textwidth}
        \centering
        \includegraphics[width=\textwidth]{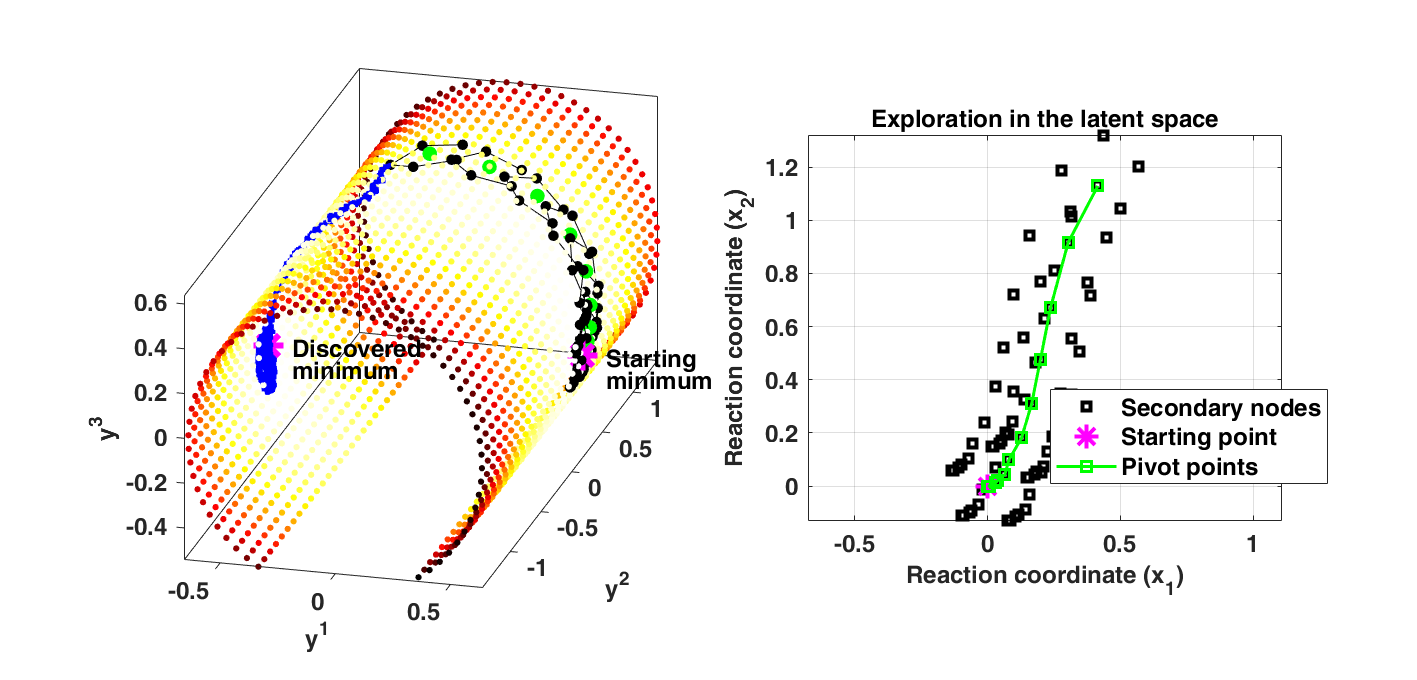}
    \end{subfigure}
    
    
    \begin{subfigure}{0.9\textwidth}
        \centering
        \includegraphics[width=\textwidth]{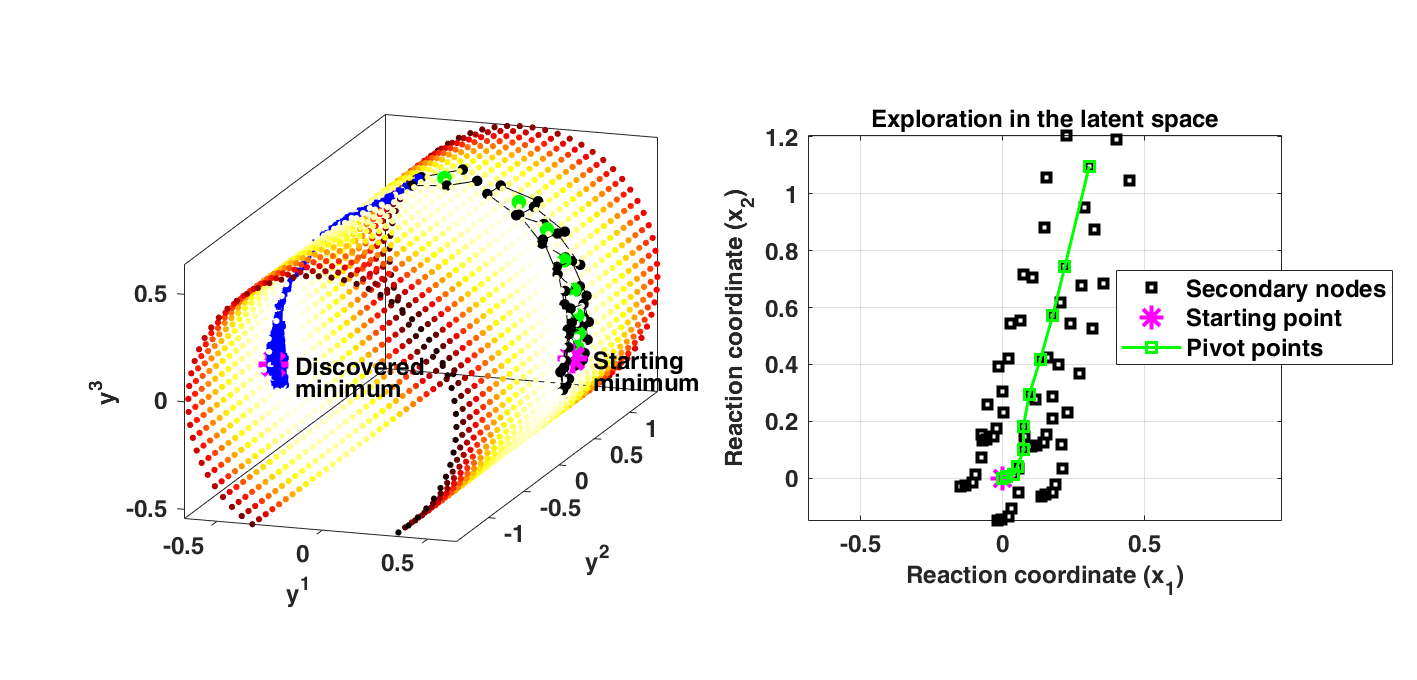}
    \end{subfigure}
    
    \caption{Active learning orchestrated search of fixed points of the SDE (\ref{SDE01}) using the ERRM procedure with $d=2$. \textbf{Top sub-panel}: The Maximum Likelihood Improvement (MLI) is used as acquisition function. \textbf{Bottom sub-panel}: The Expected Improvement (EI) is used as acquisition function. Green and black colors denote the pivot and the secondary points, respectively, in the ambient 3D space and in the 2D latent space. In both cases, one trajectory (in blue) of the SDE \ref{SDE01} starting from the furthest point discovered by the exploration in the latent space is shown to land in the unknown minimum of the potential energy.}
    \label{fig:AL:Cyl}
\end{figure}
\begin{figure}[htbp]
    \centering
    
   \begin{subfigure}{0.9\textwidth}
        \centering
        \includegraphics[width=\textwidth]{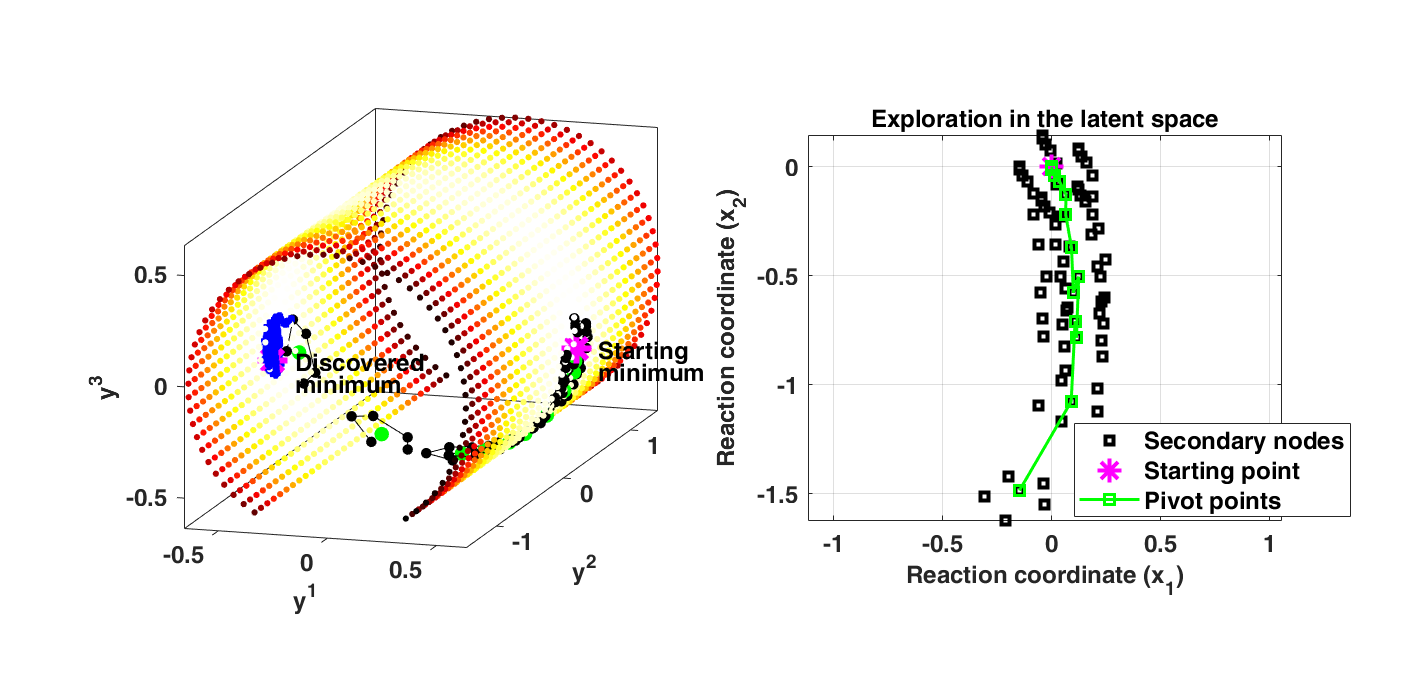}
    \end{subfigure}
    
    \caption{Active learning orchestrated search of fixed points of the SDE (\ref{SDE01}) using the explorative RRM procedure with $d=2$. The Upper Confidence Bound (UCB) is used as acqusition function with $\kappa=0.5$. Green and black colors denote the pivot and the secondary points, respectively, in the ambient 3D space and in the 2D latent space. In both cases, one trajectory (in blue) of the SDE \ref{SDE01} starting from the furthest point discovered by the exploration in the latent space is shown to land in the unknown minimum of the potential energy.}
    \label{fig:AL:Cyl01}
\end{figure}

\subsubsection{Two-dimensional implementation}\label{Cyl2Dimp}
In general, the dimension of the support of the effective energy landscape is larger than one.
Thus, it may be necessary to increase the dimension of $d$ for enhancing the stability of the RRM procedure.
Clearly, for $d>1$ the possible increase in stability comes at the cost of an increased exploration complexity.
Therefore, as discussed in the Section \ref{secnavigation}, a more sophisticated navigation strategy is nedeed.
Although not strictly necessary for the specific example (as the $d=1$ implementation is already stable), we discuss below the ERRM implementation with $d=2$ for the SDE (\ref{SDE01}).
In Figures \ref{fig:AL:Cyl} and \ref{fig:AL:Cyl01}, three successful explorations are reported where an active learning orchestration (with the three different acquisition functions reviewed in Section \ref{ALprocedure}) is utilized for navigating from one of the metastable points (i.e. the one with $y^1>0$) towards other fixed points.
Here, we use the objective function (\ref{objfun}) with: $\alpha=1, \beta=2, \epsilon=0.01$.
The relaxation step of the ERRM is applied to the pivot and the five secondary points following the same procedure described in Section \ref{Cyl1Dimp} with $D=0.5$, $N_r=10$, $q=3$, $\Delta t= 2 \times 10^{-5}$, $dx_{i,max}=0.015$, and the following initial parameterization matrix:
\begin{equation}
\textbf{D}_{in} = \begin{bmatrix}
0 & 1 & 0 \\
0.16 & 0 & 1
\end{bmatrix}
\end{equation}
The latter was computed as the null space to the local fast subspace refined at the starting pivot point: $\textbf{c}_0=[0.63,0.21;-0.10]$.
Note that the gradient $\nabla V$ in (\ref{objfun}) was estimated by averaging multiple short relaxations at the desired points.

The adopted strategy is to let the active learning orchestrator compute the correction vector by processing the pivot points only and apply the same optimal correction $[dx_{1,opt},dx_{2,opt}]$ to the selected pivot and the corresponding five secondary nodes.
Similarly to the above example in Section \ref{2DMBsec}, for the Gaussian regressor, we use an adaptive kernel parameter where, $\text{kernelScale}=0.1$ unless when between two subsequent steps the distance between the initial pivot $\textbf{c}_0$ and the corrected one $\textbf{c}_1$ is smaller than a threshold ($\| \textbf{c}_0-\textbf{c}_1 \| \le 0.05$).
In the latter case, to favor visiting unknown regions, we double the current kernel parameter until two subsequent steps deliver $\| \textbf{c}_0-\textbf{c}_1 \|>0.05$ so that the initial $\text{kernelScale}=0.1$ is restored.
It is worth noticing that in this case, in general, the path followed by the pivot points does not represent the transition path and there is no guarantee that the exploration will successfully identify a saddle point.
Possibly, with a reference to the examples reported in Figures \ref{fig:AL:Cyl} and \ref{fig:AL:Cyl01}, upon the identification of second minimum of the potential energy, the saddle point can be also found by using the string method.
\subsection{Seven-atom hexagonal Lennard Jones cluster in the plane}
Finally, we examine a seven-atom hexagonal cluster in the plane interacting via Lennard-Jones potential.
In particular, we are interested to the transition path by which the central atom moves towards the external surface \cite{dellago1998efficient,weinan2002string}.
%
\begin{figure}[htbp]
    \centering
    
    \begin{subfigure}{0.55\textwidth}
        \centering
        \includegraphics[width=\textwidth]{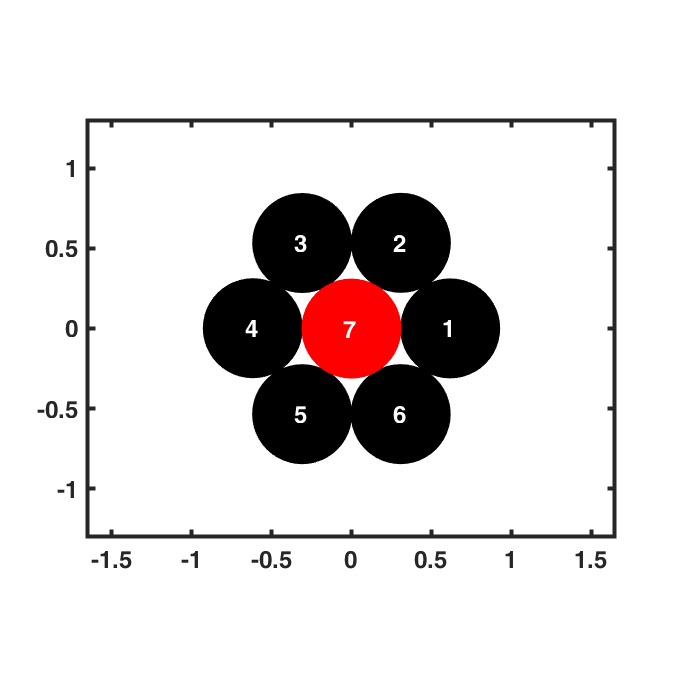}
    \end{subfigure}
    \begin{subfigure}{0.8\textwidth}
        \centering
        \includegraphics[width=\textwidth]{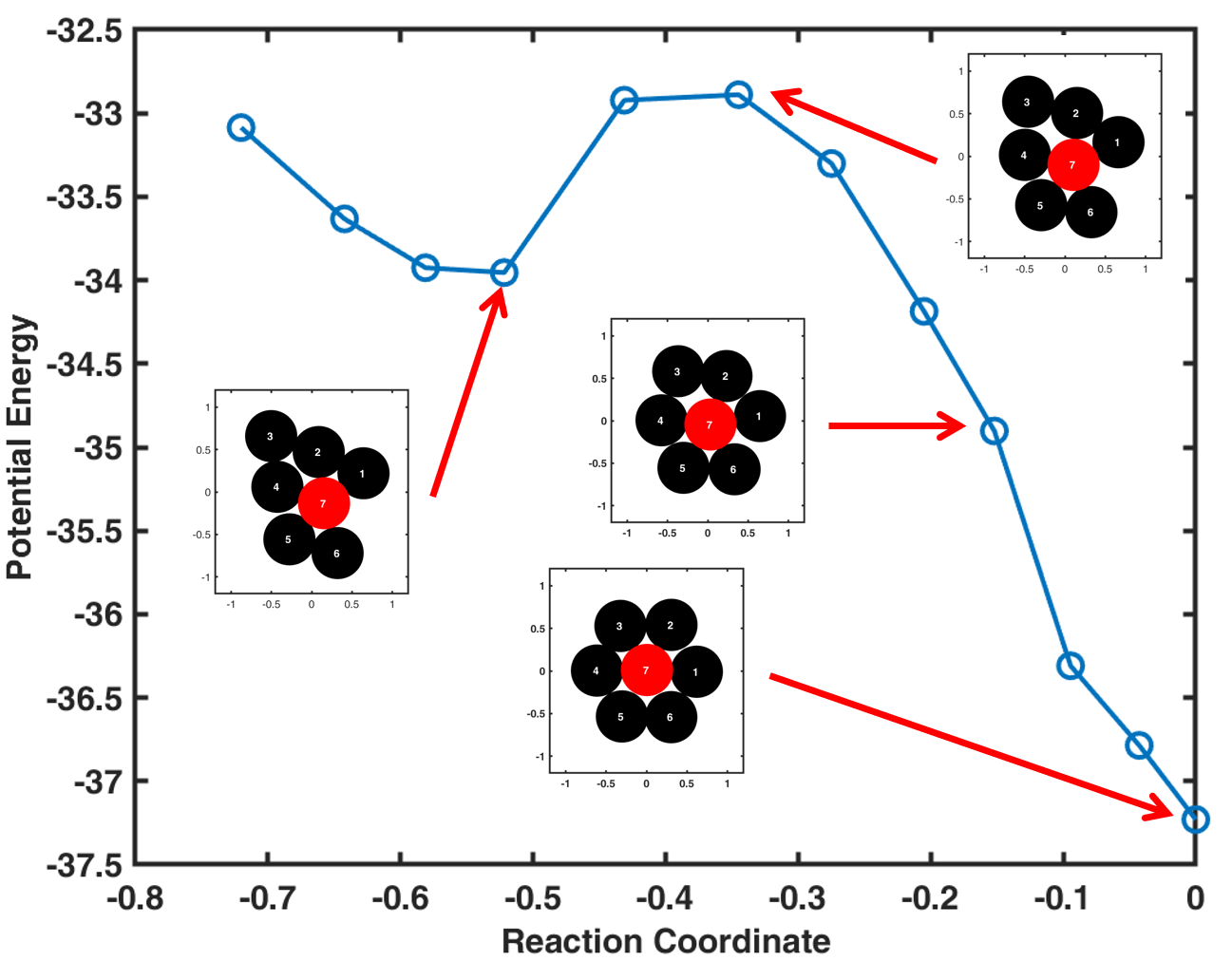}
    \end{subfigure}
        
    \caption{Top: Seven atoms interacting via Lennard-Jones potential on the plane ($\sigma=0.55$, $\epsilon=3$) with target temperature $T_{target}=0.05$, integration time step $dt=5 \times 10^{-6}$ and $\tau_T=10 \times dt$. Initially, atom 7 is centered at the origin, whereas the remaining six atoms form a surrounding hexagon. Bottom: Starting from the configuration with the lowest energy, the ERRM is applied with $d=1$ till a saddle point is reached from which the system can be relaxed to a metastable minimum. Only the pivot configurations are reported.}
    \label{fig:HA01}
\end{figure}
    
    

The system dynamics are simulated under constant temperature conditions using the Velocity Verlet integration scheme combined with a Berendsen thermostat.

The utilized simulator is treated as a black-box: Simulations are performed by assigning as an input initial positions of the particles and velocities according to a Maxwell-Boltzmann distribution (ensuring consistency with the target temperature).
Model parameters include the number of simulation steps, the Lennard-Jones potential parameters ($\sigma$) and $\epsilon$), the particle mass, the cutoff radius ($r_{cut}$) for interactions, the target temperature ($T_{target}$), and the time step ($dt$) for numerical integration.
%

The positions and velocities of the particles are updated over time using the Velocity Verlet integration scheme. 
The potential is truncated at the cutoff radius, \texttt{rcut}, to limit computational costs. 
The Lennard-Jones potential is expressed as:

\begin{equation}
V(r) = 4 \epsilon \left[ \left( \frac{\sigma}{r} \right)^{12} - \left( \frac{\sigma}{r} \right)^6 \right], \quad \text{for } r < r_\text{cut},
\end{equation}
To maintain the system at the target temperature, a Berendsen thermostat is applied periodically (every 5 time steps).
This thermostat adjusts the velocities of the particles by scaling them with a factor $\lambda$, which depends on the current and target temperatures. 
The scaling factor is computed as:
\begin{equation}
\lambda = \sqrt{1 + \frac{\Delta t}{\tau_T} \left( \frac{T_\text{target}}{T_\text{current}} - 1 \right)},
\end{equation}
where $\tau_T$ is the time constant controlling the relaxation rate toward the desired temperature.

In order to keep a symmetry-invariant representation of the system, rigid translations and rotations of the particle are constantly removed.
This is achieved by subtracting the center of mass velocity from all particles to eliminate translational motion. 
Rotational alignment is performed using the Kabsch algorithm, thus minimizing the root mean square deviation between the current configuration and the initial reference configuration.

As far as the ERRM is concerned, the navigation is conducted in a 14-dimensional space, where the generic $i-th$ dimension $y^i$ in the ambient space is the Cartesian coordinate of each atom center (made  symmetry invariant by the above rigid translation and rotation removal).
In our representation, the horizontal and vertical coordinates of atom $j$ are given by $y^{2j-1}$ and $y^{2j}$, respectively. 
In Figure \ref{fig:HA01}, we report the result of an exploration by the ERRM ($d=1$), where the initial parametrization vector is:
\begin{equation}\label{Dinhex}
\textbf{D}_{in} =
\begin{array}{cccccccccccccc}
[-0.0409 & 0.3205 & -0.4709 & -0.1647 & -0.3512 & 0.1314 & 0.4594... & \\
0.1778 & -0.0203 & 0.0025 & 0.0781 & -0.3663 & 0.3457 & -0.1011]
\end{array}
\end{equation}
Here, we used four secondary nodes $N_s=4$ (with the minimum number being two), $N_r=100$, $q=100$, $dt=5 \times 10^{-6}$, $\delta x_0=0.075$.
The secondary points are initially generated by a small perturbation of the pivot along $\textbf{D}_{in}$: $\textbf{c}_{sec}=\textbf{c}_{0} \pm 0.5  \textbf{D}_{in}$ and $\textbf{c}_{sec}=\textbf{c}_{0} \pm 0.25  \textbf{D}_{in}$.
In this case, when applying the RRM the measure of the defect of invariance (\ref{defectinvariance}) is kept below $5\%$. 
For the sake of computational efficiency, before the step (e) of Figure \ref{fig2} we also performed an unbiased short relaxation (no redistribution) of pivot and secondary points for a time of 0.05.

The estimate of $\textbf{D}_{in}$ deserves a brief discussion.
The specific values have been selected by purposely imposing a movement of atom 7 along a direction (in physical space) between atom 1 and atom 6, and by requesting that all other components in $\textbf{D}_{in}$ make it the direction (in configuration space) along which a perturbation of $\textbf{c}_0$ induces the least increase in the potential energy function.
Although only at the very beginning, this introduces a prior knowledge via such a specific choice of $\textbf{D}_{in}$.
It is therefore desirable to elaborate a more automated approach to operate such a selection in the near future.

\section{CONCLUSION and DISCUSSION}\label{conclusions}
In this work, we demonstrate that a quadratic variant of the Relaxation Redistribution Method (RRM) can play a key role in implementing an effective strategy for searching fixed points of dynamical systems.
The latter strategy is here referred to as {\it Explorative Relaxation Redistribution Method} (ERRM).
Interestingly, the ERRM does not require a prior knowledge of reaction coordinates and locations of other metastable states of the system.
Even when alternative methods available in the current literature do not assume prior knowledge of products - such as in the single-end String Method variant \cite{SingleEndString01} or in the climbing NEB \cite{climbingNEB} - the computation of energy gradients is still required and, importantly, is restricted to a one-dimensional search.
Moreover, the necessity of relying on potential energy gradients may lead to both computational challenges (associated with the energy gradient calculations) and possible detrimental neglect of entropic and dynamic effects.
On the other hand, the ERRM: i) it is intrinsically dynamic; ii) it does not require the explicit computation of the potential gradients (at least when $d=1$). 
On the contrary, in the ERRM the dynamic simulator can be treated as a black-box; iii) it does not require a prior global parametrization as it gradually discovers the latent/reaction coordinates; iv) it may attain the true intrinsic dimension of the support of the effective energy surface thus not being constrained to necessarily navigate with $d=1$ (although the latter remains the most convenient option from the computational perspective).
Interestingly, especially when the navigation is performed with $d>1$, in addition to the here suggested approaches based on active learning, recent strategies based on the {\it gentlest ascent dynamics} \cite{bello2023gentlest} or the {\it gradient extremals} \cite{georgiou2023locating} could be developed in the near future for setting up more effective navigation processes towards saddle points.

Open questions to be addressed in the near future still remain as briefly described below.
First, the initial parametrization matrix $\textbf{D}_{in}$ can be selected by minimizing the energy increase from the starting configuration, but clearly a more automated selection has to be developed especially when $d>1$.
Furthermore, an automated choice of RRM parameters (e.g. the initially chosen spacing between the pivot and the secondary points, the duration of the relaxation step before a redistribution step is performed, the number $N_r$ of relaxations per point) is desirable and shall be addressed in future works.
Similarly, for cases with $d>1$, a comprehensive study for devising more general and effective objective functions alternative to (\ref{objfun}) are desirable.

A critical aspect for the successful implementation of the method is the possibility of having access to a fully symmetry-invariant representation of the system of interest, which often comes in terms of Cartesian coordinates.
Needless to say that the latter aspect may reveal particularly challenging, especially when dealing with more complex systems.
In this respect, the use of modern frameworks to achieve translation-, rotation- and permutation-free descriptions appear particularly appealing \cite{ACSF,soap2016}.

Finally, new interesting research avenues may also emerge by the adoption of the recently suggested methods for automatic discovery of good reduced variables \cite{barletta2024learning,trezza2023leveraging} applied to the data collected during the ERRM exploration.

\section*{Acknowledgments}
We acknowledge funding under the National Recovery and 957 Resilience Plan (NRRP), Mission 4 Component 2 Investment 1.3-Call for tender No. 1561 of 11.10.2022 of Ministero dell'Università e della Ricerca (MUR); funded by the European Union NextGenerationEU.

\subsection*{Data Availability}
Data and codes are available upon reasonable request to the corresponding Author.


\begin{thebibliography}{10}
\expandafter\ifx\csname url\endcsname\relax
  \def\url#1{\texttt{#1}}\fi
\expandafter\ifx\csname urlprefix\endcsname\relax\def\urlprefix{URL }\fi
\expandafter\ifx\csname href\endcsname\relax
  \def\href#1#2{#2} \def\path#1{#1}\fi

\bibitem{dill2008protein}
K.~A. Dill, S.~B. Ozkan, M.~S. Shell, T.~R. Weikl, The protein folding problem, Annu. Rev. Biophys. 37~(1) (2008) 289--316.

\bibitem{bernardi2015enhanced}
R.~C. Bernardi, M.~C. Melo, K.~Schulten, Enhanced sampling techniques in molecular dynamics simulations of biological systems, Biochimica et Biophysica Acta (BBA)-General Subjects 1850~(5) (2015) 872--877.

\bibitem{hummer2003coarse}
G.~Hummer, I.~G. Kevrekidis, Coarse molecular dynamics of a peptide fragment: Free energy, kinetics, and long-time dynamics computations, The Journal of chemical physics 118~(23) (2003) 10762--10773.

\bibitem{noe2009constructing}
F.~No{\'e}, C.~Sch{\"u}tte, E.~Vanden-Eijnden, L.~Reich, T.~R. Weikl, Constructing the equilibrium ensemble of folding pathways from short off-equilibrium simulations, Proceedings of the National Academy of Sciences 106~(45) (2009) 19011--19016.

\bibitem{faccioliprotein}
P.~Faccioli, M.~Sega, F.~Pederiva, H.~Orland, Dominant pathways in protein folding, Physical review letters 97~(10) (2006) 108101.

\bibitem{anslyn2006modern}
E.~Anslyn, Modern physical organic chemistry (2006).

\bibitem{raucci2022discover}
U.~Raucci, V.~Rizzi, M.~Parrinello, Discover, sample, and refine: Exploring chemistry with enhanced sampling techniques, The Journal of Physical Chemistry Letters 13~(6) (2022) 1424--1430.

\bibitem{torrie1977nonphysical}
G.~M. Torrie, J.~P. Valleau, Nonphysical sampling distributions in monte carlo free-energy estimation: Umbrella sampling, Journal of computational physics 23~(2) (1977) 187--199.

\bibitem{sugita1999replica}
Y.~Sugita, Y.~Okamoto, Replica-exchange molecular dynamics method for protein folding, Chemical physics letters 314~(1-2) (1999) 141--151.

\bibitem{zhou2006replica}
R.~Zhou, Replica exchange molecular dynamics method for protein folding simulation, Protein Folding Protocols (2006) 205--223.

\bibitem{yang2019enhanced}
Y.~I. Yang, Q.~Shao, J.~Zhang, L.~Yang, Y.~Q. Gao, Enhanced sampling in molecular dynamics, The Journal of chemical physics 151~(7) (2019).

\bibitem{kang2024computing}
P.~Kang, E.~Trizio, M.~Parrinello, Computing the committor with the committor: an anatomy of the transition state ensemble, arXiv preprint arXiv:2401.05279 (2024).

\bibitem{NEB01}
H.~J{\'o}nsson, G.~Mills, K.~W. Jacobsen, Nudged elastic band method for finding minimum energy paths of transitions, in: Classical and quantum dynamics in condensed phase simulations, World Scientific, 1998, pp. 385--404.

\bibitem{weinan2002string}
E.~Weinan, W.~Ren, E.~Vanden-Eijnden, String method for the study of rare events, Physical Review B 66~(5) (2002) 052301.

\bibitem{Meta01}
A.~Laio, M.~Parrinello, Escaping free-energy minima, Proceedings of the national academy of sciences 99~(20) (2002) 12562--12566.

\bibitem{bonati2021deep}
L.~Bonati, G.~Piccini, M.~Parrinello, Deep learning the slow modes for rare events sampling, Proceedings of the National Academy of Sciences 118~(44) (2021) e2113533118.

\bibitem{jung2023machine}
H.~Jung, R.~Covino, A.~Arjun, C.~Leitold, C.~Dellago, P.~G. Bolhuis, G.~Hummer, Machine-guided path sampling to discover mechanisms of molecular self-organization, Nature Computational Science 3~(4) (2023) 334--345.

\bibitem{chiavazzo2017PNAS}
E.~Chiavazzo, R.~Covino, R.~R. Coifman, C.~W. Gear, A.~S. Georgiou, G.~Hummer, I.~G. Kevrekidis, Intrinsic map dynamics exploration for uncharted effective free-energy landscapes, Proceedings of the National Academy of Sciences 114~(28) (2017) E5494--E5503.

\bibitem{georgiou2017exploration}
A.~S. Georgiou, J.~M. Bello-Rivas, C.~W. Gear, H.-T. Wu, E.~Chiavazzo, I.~G. Kevrekidis, An exploration algorithm for stochastic simulators driven by energy gradients, Entropy 19~(7) (2017) 294.

\bibitem{faccioliiMapD}
D.~Ghamari, R.~Covino, P.~Faccioli, Sampling a rare protein transition using quantum annealing, Journal of Chemical Theory and Computation 20~(8) (2024) 3322--3334.

\bibitem{Clementi1d}
W.~Zheng, M.~A. Rohrdanz, C.~Clementi, Rapid exploration of configuration space with diffusion-map-directed molecular dynamics, The journal of physical chemistry B 117~(42) (2013) 12769--12776.

\bibitem{Maggioni2011}
M.~A. Rohrdanz, W.~Zheng, M.~Maggioni, C.~Clementi, Determination of reaction coordinates via locally scaled diffusion map, The Journal of chemical physics 134~(12) (2011).

\bibitem{bello2023gentlest}
J.~M. Bello-Rivas, A.~Georgiou, H.~Vandecasteele, I.~G. Kevrekidis, Gentlest ascent dynamics on manifolds defined by adaptively sampled point-clouds, The Journal of Physical Chemistry B 127~(23) (2023) 5178--5189.

\bibitem{coifman2005geometric}
R.~R. Coifman, S.~Lafon, A.~B. Lee, M.~Maggioni, B.~Nadler, F.~Warner, S.~W. Zucker, Geometric diffusions as a tool for harmonic analysis and structure definition of data: Diffusion maps, Proceedings of the national academy of sciences 102~(21) (2005) 7426--7431.

\bibitem{chiavazzo2014processes}
E.~Chiavazzo, C.~W. Gear, C.~J. Dsilva, N.~Rabin, I.~G. Kevrekidis, Reduced models in chemical kinetics via nonlinear data-mining, Processes 2~(1) (2014) 112--140.

\bibitem{chiavazzo2011adaptive}
E.~Chiavazzo, I.~Karlin, Adaptive simplification of complex multiscale systems, Physical Review E—Statistical, Nonlinear, and Soft Matter Physics 83~(3) (2011) 036706.

\bibitem{chiavazzo2009PhDThesis}
E.~Chiavazzo, Invariant manifolds and lattice boltzmann method for combustion, Ph.D. thesis, ETH Zurich (2009).

\bibitem{chiavazzo2012approximation}
E.~Chiavazzo, Approximation of slow and fast dynamics in multiscale dynamical systems by the linearized relaxation redistribution method, Journal of Computational Physics 231~(4) (2012) 1751--1765.

\bibitem{kooshkbaghi2014global}
M.~Kooshkbaghi, C.~E. Frouzakis, E.~Chiavazzo, K.~Boulouchos, I.~V. Karlin, The global relaxation redistribution method for reduction of combustion kinetics, The Journal of chemical physics 141~(4) (2014).

\bibitem{maragliano2006string}
L.~Maragliano, A.~Fischer, E.~Vanden-Eijnden, G.~Ciccotti, String method in collective variables: Minimum free energy paths and isocommittor surfaces, The Journal of chemical physics 125~(2) (2006).

\bibitem{frewen2009exploration}
T.~A. Frewen, G.~Hummer, I.~G. Kevrekidis, Exploration of effective potential landscapes using coarse reverse integration, The Journal of chemical physics 131~(13) (2009).

\bibitem{gear2003equation}
C.~W. Gear, J.~M. Hyman, P.~G. Kevrekidid, I.~G. Kevrekidis, O.~Runborg, C.~Theodoropoulos, Equation-free, coarse-grained multiscale computation: Enabling mocroscopic simulators to perform system-level analysis (2003).

\bibitem{kevrekidis2004equation}
I.~G. Kevrekidis, C.~W. Gear, G.~Hummer, Equation-free: The computer-aided analysis of complex multiscale systems, AIChE Journal 50~(7) (2004) 1346--1355.

\bibitem{DirkPaper}
V.~Reinhardt, M.~Winckler, D.~Lebiedz, Approximation of slow attracting manifolds in chemical kinetics by trajectory-based optimization approaches, The Journal of Physical Chemistry A 112~(8) (2008) 1712--1718.

\bibitem{lebiedz2004computing}
D.~Lebiedz, Computing minimal entropy production trajectories: An approach to model reduction in chemical kinetics, The Journal of chemical physics 120~(15) (2004) 6890--6897.

\bibitem{ILDM}
U.~Maas, S.~B. Pope, Simplifying chemical kinetics: intrinsic low-dimensional manifolds in composition space, Combustion and flame 88~(3-4) (1992) 239--264.

\bibitem{ValoraniPaper}
M.~Valorani, D.~A. Goussis, F.~Creta, H.~N. Najm, Higher order corrections in the approximation of low-dimensional manifolds and the construction of simplified problems with the csp method, Journal of Computational Physics 209~(2) (2005) 754--786.

\bibitem{GorbanBook}
A.~N. Gorban, I.~V. Karlin, Invariant manifolds for physical and chemical kinetics, Vol. 660, Springer, 2005.

\bibitem{chiavazzo2007comparison}
E.~Chiavazzo, A.~N. Gorban, I.~V. Karlin, et~al., Comparison of invariant manifolds for model reduction in chemical kinetics, Commun. Comput. Phys 2~(5) (2007) 964--992.

\bibitem{bonke2024multi}
S.~A. Bonke, G.~Trezza, L.~Bergamasco, H.~Song, S.~Rodriguez-Jimenez, L.~Hammarstroem, E.~Chiavazzo, E.~Reisner, Multi-variable multi-metric optimization of self-assembled photocatalytic co2 reduction performance using machine learning algorithms, Journal of the American Chemical Society (2024).

\bibitem{trezza2022minimal}
G.~Trezza, L.~Bergamasco, M.~Fasano, E.~Chiavazzo, Minimal crystallographic descriptors of sorption properties in hypothetical mofs and role in sequential learning optimization, npj Computational Materials 8~(1) (2022) 123.

\bibitem{isomap2002}
M.~Balasubramanian, E.~L. Schwartz, The isomap algorithm and topological stability, Science 295~(5552) (2002) 7--7.

\bibitem{ling2017high}
J.~Ling, M.~Hutchinson, E.~Antono, S.~Paradiso, B.~Meredig, High-dimensional materials and process optimization using data-driven experimental design with well-calibrated uncertainty estimates, Integrating Materials and Manufacturing Innovation 6 (2017) 207--217.

\bibitem{dellago1998efficient}
C.~Dellago, P.~G. Bolhuis, D.~Chandler, Efficient transition path sampling: Application to lennard-jones cluster rearrangements, The Journal of chemical physics 108~(22) (1998) 9236--9245.

\bibitem{SingleEndString01}
P.~M. Zimmerman, Single-ended transition state finding with the growing string method, Journal of computational chemistry 36~(9) (2015) 601--611.

\bibitem{climbingNEB}
G.~Henkelman, B.~P. Uberuaga, H.~J{\'o}nsson, A climbing image nudged elastic band method for finding saddle points and minimum energy paths, The Journal of chemical physics 113~(22) (2000) 9901--9904.

\bibitem{georgiou2023locating}
A.~Georgiou, H.~Vandecasteele, J.~Bello-Rivas, I.~Kevrekidis, Locating saddle points using gradient extremals on manifolds adaptively revealed as point clouds, Chaos: An Interdisciplinary Journal of Nonlinear Science 33~(12) (2023).

\bibitem{ACSF}
J.~Behler, Atom-centered symmetry functions for constructing high-dimensional neural network potentials, The Journal of chemical physics 134~(7) (2011).

\bibitem{soap2016}
S.~De, A.~P. Bart{\'o}k, G.~Cs{\'a}nyi, M.~Ceriotti, Comparing molecules and solids across structural and alchemical space, Physical Chemistry Chemical Physics 18~(20) (2016) 13754--13769.

\bibitem{barletta2024learning}
G.~Barletta, G.~Trezza, E.~Chiavazzo, Learning effective good variables from physical data, Machine Learning and Knowledge Extraction 6~(3) (2024) 1597--1618.

\bibitem{trezza2023leveraging}
G.~Trezza, E.~Chiavazzo, Leveraging composition-based energy material descriptors for machine learning models, Materials Today Communications 36 (2023) 106579.

\end{thebibliography}

\end{document}